\documentclass[12pt,manuscript]{emulateapj}

\usepackage{amsmath}

\newcommand\T{\rule{0pt}{2.6ex}}
\newcommand\B{\rule{-1.2ex}{0pt}}

\slugcomment{Accepted for publication in the Astrophysical Journal}

\shorttitle{BLAST Number Counts}
\shortauthors{Patanchon et al. 2009}

\begin{document}

\title{Submillimeter Number Counts From Statistical Analysis of BLAST Maps}

\author{
Guillaume~Patanchon,\altaffilmark{1,\dag}
Peter~A.~R.~Ade,\altaffilmark{2}
James~J.~Bock,\altaffilmark{3}
Edward~L.~Chapin,\altaffilmark{4}
Mark~J.~Devlin,\altaffilmark{5}
Simon~R.~Dicker,\altaffilmark{5}
Matthew~Griffin,\altaffilmark{2}
Joshua~O.~Gundersen,\altaffilmark{6}
Mark~Halpern,\altaffilmark{4}
Peter~C.~Hargrave,\altaffilmark{2}
David~H.~Hughes,\altaffilmark{7}
Jeff~Klein,\altaffilmark{5}
Gaelen~Marsden,\altaffilmark{4}
Philip~Mauskopf,\altaffilmark{2}
Lorenzo~Moncelsi,\altaffilmark{2}
Calvin~B.~Netterfield,\altaffilmark{8,9}
Luca~Olmi,\altaffilmark{10,11}
Enzo~Pascale,\altaffilmark{2}
Marie~Rex,\altaffilmark{5}
Douglas~Scott,\altaffilmark{4}
Christopher~Semisch,\altaffilmark{5}
Nicholas~Thomas,\altaffilmark{6}
Matthew~D.~P.~Truch,\altaffilmark{5}
Carole~Tucker,\altaffilmark{2}
Gregory~S.~Tucker,\altaffilmark{12}
Marco~P.~Viero,\altaffilmark{8}
Donald~V.~Wiebe\altaffilmark{4,9} 
}

\altaffiltext{1}{Universit{\'e} Paris Diderot, Laboratoire APC, 10,
  rue Alice Domon et L{\'e}onie Duquet 75205 Paris, France.}

\altaffiltext{2}{Department of Physics \& Astronomy, University of
  British Columbia, 6224 Agricultural Road, Vancouver, BC V6T~1Z1,
  Canada}

\altaffiltext{3}{School of Physics \& Astronomy, Cardiff University, 5
  The Parade, Cardiff, CF24 3AA, UK.}

\altaffiltext{4}{Jet Propulsion Laboratory, Pasadena, CA 91109-8099,
  USA.}

\altaffiltext{5}{Department of Physics \& Astronomy, University of
  Pennsylvania, 209 South 33rd Street, Philadelphia, PA, 19104, USA.}

\altaffiltext{6}{Department of Physics, University of Miami, 1320
  Campo Sano Drive, Coral Gables, FL 33146, USA.}

\altaffiltext{7}{Instituto Nacional de Astrof\'isica \'Optica y
  Electr\'onica (INAOE), Aptdo. Postal 51 y 72000 Puebla, Mexico.}

\altaffiltext{8}{Department of Astronomy \& Astrophysics, University
  of Toronto, 50 St. George Street Toronto, ON M5S~3H4, Canada.}

\altaffiltext{9}{Department of Physics, University of Toronto, 60
  St. George Street, Toronto, ON M5S~1A7, Canada.}

\altaffiltext{10}{University of Puerto Rico, Rio Piedras Campus, Physics Dept.,
 Box 23343, UPR station, Puerto Rico 00931.}

\altaffiltext{11}{INAF, Osservatorio Astrofisico di Arcetri, Largo
  E. Fermi 5, I-50125, Firenze, Italy}

\altaffiltext{12}{Department of Physics, Brown University, 182 Hope
  Street, Providence, RI 02912, USA.}

\altaffiltext{\dag}{\url{patanchon@apc.univ-paris-diderot.fr}}

\begin{abstract}

  We describe the application of a statistical method to estimate
  submillimeter galaxy number counts from confusion limited
  observations by the Balloon-borne Large Aperture Submillimeter
  Telescope (BLAST).  Our method is based on a maximum likelihood fit
  to the pixel histogram, sometimes called `$P(D)$', an approach which
  has been used before to probe faint counts, the difference being
  that here we advocate its use even for sources with relatively high
  signal-to-noise ratios.  This method has an advantage over standard
  techniques of source extraction in providing an unbiased estimate of
  the counts from the bright end down to flux densities well below the
  confusion limit.  We specifically analyse BLAST observations of a
  roughly $10\,{\rm deg}^2$ map centered on the Great Observatories
  Origins Deep Survey South field.  We provide estimates of number
  counts at the three BLAST wavelengths, 250, 350, and 500\,\micron;
  instead of counting sources in flux bins we estimate the counts at
  several flux density nodes connected with power-laws.  We observe a
  generally very steep slope for the counts of about $-3.7$ at
  250$\,\mu$m and $-4.5$ at 350 and 500$\,\mu$m, over the range
  $\sim $0.02--0.5$\,$Jy, breaking to a shallower slope below about
  0.015$\,$Jy at all three wavelengths.  We also
  describe how to estimate the uncertainties and correlations in this
  method so that the results can be used for model-fitting.  This
  method should be well-suited for analysis of data from the {\em
    Herschel\/} satellite.

\end{abstract}


\hyphenation{Sub-milli-meter}
\hyphenation{sub-milli-meter}

\section{Introduction}

When the very first surveys are taken in any wavelength band, counting
the number of sources found as a function of source apparent
brightness is an efficient method for learning about the population of
sources uncovered. Typically this approach provides clues much more
rapidly than the painstaking work of identifying and studying the
sources individually.  If the sources lie nearby on a cosmic scale,
one expects the number of sources per unit solid angle brighter than
some limiting flux density $S$ to vary as $N(>S)\propto S^{-3/2}$, the
Euclidean limit.  Evolution in the volume density of sources, or their
luminosity over time, causes a departure from the Euclidean slope,
such that counts measurements can be used to infer information about
the history of the population.  In one of the first quantitative
applications of this technique, Eddington wrote in {\it The Large
  Scale Structure of the Universe} (1911) that the Universe consists
of $10^{10} $ or $10^{11}$ stars surrounded by vast amounts of empty
space.  Clearly a large spatial inhomogeneity also generates
non-Euclidean features, and this was a correct inference given that
galaxies had not yet been discovered.  On a cosmic scale we do not
anticipate discovering such large inhomogeneities, but genuine
clustering of sources and {\it cosmic variance}, where different
regions happen to have different densities, both of which will effect
measured source counts.

At $24~\mu$m source counts follow a Euclidean distribution until just
near the faintest end of the deepest surveys
\citep{Shupe08,Papovich03,Rodighero06,Chary04,Marleau04}, implying a
fairly uniformly distributed stable population at the bright end.
However, the $850\,\mu$m-selected sources in SCUBA surveys follow a
very steep broken power law distribution (see \citealt{Coppin06,Knudsen08,Smail02,Scott02,Webb03,Borys03}; and
also \citealt{Austermann09a} at mm wavelengths).  At $850\,\mu$m there is enough of a negative $k-$correction that
there is little variation in the apparent brightness of a source with
a given luminosity at redshifts $1<z<8$. Furthermore, the increasing
volume sampled at higher redshifts enables surveys at this wavelength
to efficiently sample large numbers of distant objects. The
contrasting shapes of the counts distributions at 24 and 850$\,\mu$m
implies both that the brightest sources are rare, and that their
numbers have decreased over time (\citealt{Papovich03}; for measurements at intermediate wavelengths see \citealt{Frayer09,Dole04}).  We report here
surveys made with BLAST, the first statistically useful surveys in the
crucial spectral range from $200\,\mu$m to $600\,\mu$m, near the peak of
the Cosmic Infrared Background \citep{Puget96,Fixsen98}.  These
surveys will explore the transition between the nearby luminous
galaxies and the distant starburst population.

Counting objects in the sky is an endeavor which is probably as old as
counting itself.  In astronomy determining the abundance of objects as
a function of apparent brightness is often the easiest way to describe
a population, since detailed spectral information is usually required
in order to extract intrinsic properties of objects.  Hence a great
deal has been written about how to estimate `number counts'
efficiently.  The process includes carrying out estimates of
incompleteness, flux boosting and corrections for other sources of
bias.

Radio astronomers discovered in the 1950s \citep{Scheuer57} that one
could use the statistical properties of observations of the sky to
probe the counts of sources which are too faint to detect individually
\citep[see also][]{Murdoch73,Scheuer74,Condon74,Barcons92,Takeuchi04}.
The `probability of deflection' or $P(D)$ distribution is essentially
the histogram of pixel values in a map, and it depends on the
underlying source counts.  For simple distributions, particularly
power-law counts, it is relatively easy to estimate the amplitude and
slope of the confused source counts.  The conventional approach has
been to count brighter objects directly and to carry out a $P(D)$
analysis at the faint end.  However, we have found that, at least in
regimes where a flux boosting bias is important \citep{Coppin05}, it
is better to use a histogram-fitting procedure for the full range of
source brightnesses.  In other words if one wants to obtain a robust
estimate of the source counts, it is better to avoid counting any
objects at all, a somewhat counterintuitive result.

This paper specifically examines data from the Balloon-borne Large
Aperture Submillimeter Telescope (BLAST) at 250, 350 and 500\,\micron.
A first estimate of the counts at all three BLAST wavelengths was
presented in \citet{Devlin09}.  Here we present the method in much
more detail, and perform a refined analysis including a comprehensive
discussion of uncertainties and a discussion of clustering.

Our early attempts to estimate BLAST source counts relied on the
traditional approach of thresholding the maps in signal-to-noise ratio
(S/N), extracting candidate sources, and then estimating corrections
for `flux-boosting', reliability, incompleteness, etc. This approach
did not yield useful results, even for ${\rm S/N}\,{\geq}\,5$ sources.
The BLAST data that we examined consist of a 2-tier survey, with a
smaller region having much deeper integration than the bulk of the map
area.  In practice we find it very difficult to match the counts for
the range of flux densities where the 2 tiers overlap. Application of
the method to simulated data-sets reveals strong biases in the
estimated counts.  We are therefore led to pursue other approaches,
motivated additionally by earlier attempts to study the $P(D)$
statistics of data from the SCUBA instrument, finding that a careful
modeling of the counts to fit the pixel histogram achieves much more
satisfactory results.

We emphasize that in the S/N regime probed with BLAST, and future
surveys such as those that will be undertaken by the {\em Herschel\/}
satellite, the statistical fitting of the pixel histogram gives better
results than standard techniques of source extraction over the full
flux range.

There have been many previous $P(D)$-style studies of source counts
\citep[e.g.][]{Franceschini89,WallSPW82,BarconsF90,Oliver97,Maloney05},
but typically they were restricted to studying the faint end of the
counts, and using a single power-law for the underlying model.  The
closest study to our own in the literature is by \citet{Friedmann04}.
Those authors developed a minimum $\chi^2$ approach and applied it to
simulations of {\em ISO\/} data from the FIRBACK survey
\citep{Puget99} to fit a double power-law model.  In this paper, we
have pursued this approach and have developed a maximum likelihood
method applied to data of significantly higher quality and quantity.
We have uncovered a number of issues related to application to real
data. We have also developed a more efficient implementation of
several steps in the analysis, as well as techniques to accurately
estimate errors.  Discussion of these details is likely to be helpful
when applying a similar approach to even better data-sets. Our method
accounts for issues related to realistic instrumental noise and
pre-processing of the map. We provide solutions to deal with
inhomogeneous and large scale noise in the map, and to correct the
effect of optional map filtering. We also discuss in detail the choice
of filter to apply to the maps for the optimization of source count
estimation.

We present the application of the method to multi-power-law count
models using Markov Chains \citep[e.g.][]{Chib95} to sample the likelihood and provide an
extended discussion of uncertainties and correlation among
parameters. We discuss how to marginalize over the total background
intensity, a quantity which is not accessible from the data and
examine how to include prior information on the background in the
analysis.

Our paper is organized as follows: we introduce the BLAST data in
\S~\ref{sec:BLASTobs}, we present the model of observations and the
main steps of the derivation of the probability function in
\S~\ref{sec:Model}. The maximum likelihood method is developed in
\S~\ref{sec:method}, and in \S~\ref{sec:Appli} we present the
application to BLAST maps and provide estimates of the counts at 250,
350, and 500$\,\micron$. The comparison with other data and extensions
of the method are discussed in \S~\ref{sec:Discuss}
and~\ref{sec:conclusion}.

\section{BLAST observations}
\label{sec:BLASTobs}

BLAST is a stratospheric balloon-borne telescope incorporating a 1.8-m
primary mirror, and operating at an altitude of approximately 39\,km.
The focal plane is populated with three bolometer arrays observing in
contiguous bands with central wavelengths of 250, 350, and
500$\,\mu$m, essentially a prototype of the camera of the Spectral and
Photometric Imaging Receiver (SPIRE) for {\em Herschel\/}
\citep{Griffin07}.  BLAST had two successful scientific flights.  Here
we use data from the 11-day flight carried out in 2006 from McMurdo
Station, Antarctica. The under-illuminated BLAST primary produced
nearly diffraction-limited beams with full-width at half-maxima (FWHM)
of 36$\arcsec$ 42$\arcsec$, and 60$\arcsec$, at 250, 350, and
$500\,\mu$m, respectively.  BLAST deep and wide blank-field surveys,
hereafter BGS-Deep and BGS-Wide, were centered on the Great
Observatories Origins Deep Survey South (GOODS-S) field, which is at
the centre of the Chandra Deep Field South (CDFS).  A second
intermediate depth field near the South Ecliptic Pole was also
surveyed. In addition, BLAST also targeted parts of the Milky Way
\citep{Netterfield09} and some nearby galaxies. Several other
observations of low-redshift clusters and high-redshift radio galaxies
were made to sample biased star-forming regions of the
Universe.\footnote{See {\tt http://blastexperiment.info} for more
  details.}  Further details of the instrument can be found in
\citet{Pascale08}, and the flight performance and calibration for the
2006 flight are provided in \citet{Truch09}. In this paper we focus on
the BGS-Deep+Wide map, which covers an area of approximately
10$\,\rm{deg}^2$. The deep part, nested inside the wide, has an area
of 0.8$\,\rm{deg}^2$ and is confusion limited in all three bands in
the sense that the variance of the map at the scale of the point
spread function is dominated by sources rather than noise.

The processing of BLAST timestreams includes despiking, correcting
time-varying detector responsivities, and deconvolving the effects of
detector thermal time constants and audio frequency filtering. The
absolute calibration is based on regular observations of the evolved
star VY CMa, which results in systematic uncertainties common to the
three BLAST bands of approximately 10$\%$ \citep{Truch09}. The
calibration uncertainty propagates directly to our counts estimates,
and we neglect it from here on, since it only affects the comparison
between our results and those of other experiments.  Pointing is
reconstructed to an accuracy ${<}\,5\arcsec$.  Maps are produced using
SANEPIC, a maximum likelihood method \citep{Patanchon08} dealing with low
frequency noise as well as noise correlations between detectors.
Because of the modest scanning angle variations of this particular
field, residual correlated noise is still present at large angular
scales after map-making. A relatively weak high-pass filter is
therefore applied to the maps, suppressing signal on scales larger
than about 8$\arcmin$. The filter is anisotropic and stronger in the
main cross-scan direction. Filtering has little impact on point
sources and is accounted for in the analysis (see
\S~\ref{sub:filtpart}).

\section{Model of the observations}
\label{sec:Model}

In this section, we present the main steps of the computation of the
probability distribution function (PDF) with respect to models of
submillimeter galaxy number counts with typical observational
parameters.  Detailed derivations and descriptions of the statistics
of source confusion can be found in several articles \citep[e.g.][and
references therein]{Takeuchi01}.\footnote{For an introduction to this
  topic we recommend the Appendix of \citet{WallSPW82} or \S~2 of
  \citet{Takeuchi01}; see also \S~9.1 of \citet{Trimble05}.}.  Here we
give only a brief overview of the statistics of the `$P(D)$'
histogram.

\subsection{The probability of deflection}

Let us define $n(S)$ to be the differential number counts, i.e., the
derivative of the cumulative source counts:
\begin{equation}
  n(S) \equiv -{dN({>}S) \over dS},
\end{equation}
where $N({>}S)$ is the total number of sources per unit solid angle
with flux densities larger than $S$. Let us assume that we perform an
observation at a random position $\bold{r_0}$ on the sky. A point
source at position $\bold{r}$ is observed with flux $x = S\times
f(\bold{r}-\bold{r_0})$, where $f(\bold{\Delta r})$ is the beam
function\footnote{In a pixelised map, $f(\bold{\Delta r})$ is
  basically the result of the convolution of the experimental beam
  function with the pixel window function.}.  Then, the mean number
density of sources observed with a flux $x$ is given by the well known
result \citep{Condon74}:
\begin{equation}
  R(x) = \int n\left({x\over f(\bold{r}-\bold{r_0})}\right)
 {d^2\bold{r} \over f(\bold{r}-\bold{r_0})}.
\label{eq:obsflux}
\end{equation}
Now let $m_k$ be the total number of sources observed with a flux
between $x_k$ and $x_k + \Delta x$, and $\overline{m}_k$ the expected
number of sources in the same flux bin. Assuming that $\Delta x \ll
x_k$, we can write:
\begin{equation}
  \overline{m}_k = R(x_k) \Delta x.
\end{equation}

For this paper we assume that the sources are randomly distributed
over the sky without spatial correlations on scales larger than a beam
size.  This assumption is discussed in \S~\ref{sub:scClustering}). In
observed flux bin $k$, the probability distribution of $m_k$ follows
Poisson statistics with mean $\overline{m}_k$,
\begin{equation}
  p_k(m_k) = {(R(x_k)\Delta x)^{m_k} \over m_k!} e^{-R(x_k)\Delta x}.
\end{equation}
We can also write the characteristic function, i.e., the Fourier
transform of the probability distribution of the Poisson distribution,
as follows:
\begin{equation}
  {\tilde p_k}(\tau) = \exp[R(x_k)\Delta x(e^{i\tau}-1)].
\label{eq:characfunct}
\end{equation}

The total flux, $d_S$, from all the sources in the pixel can be
written as a sum over all the flux bins, i.e.
\begin{equation}
  d_S = \sum_{k=0}^{\infty} s_k = \sum_{k=0}^{\infty} x_k~m_k,
\label{eq:totalflux}
\end{equation}
where $s_k$ is the total flux due to all sources with observed flux
$x_k = k \Delta x$.  We want to obtain the PDF of $d_S$. Given that
the probability distribution of $d_s$ is the result of convolution of
the probability distributions of $s_k$, then individual characteristic
functions multiply. We can thus obtain from equations
(\ref{eq:characfunct}) and (\ref{eq:totalflux}) the expression for the
probability distribution of $d_S$:
\begin{equation}
\begin{split}
  p(d_S)={\cal{F}}^{-1}\left\{\exp\left(\int\limits_{0}^{\infty}R(x)\,e^{iwx}\,
\mathrm{d}x\right.\right.\\\left.\left.- \int\limits_{0}^{\infty}R(x)\,\mathrm{d}x\right)\right\}(d_S).
\label{eq:pdf}
\end{split}
\end{equation}
Here we have taken the limit of the sum in
equation~(\ref{eq:totalflux}) as an integral over $dx$, and $w$ is the
variable of the Fourier transform of $p(d_S)$.

For most experiments, and explicitly for BLAST, the mean value of the
flux density is not accessible to observation. The measured signal,
$d$, is the sum of the total response to the source flux, $d_S$, an
offset $\mu$, which we assume is constant for all pixels, and
instrumental and photon noise (which we will hereafter refer to as simply
`noise'). Assuming the noise is Gaussian
with standard deviation $\sigma_{\rm n}$, the probability distribution
of the observed signal is the convolution of the probability
distribution of pure signal and of noise, which gives:
\begin{equation}
\begin{split}
  p(d) = {\cal{F}}^{-1}\left\{
  \exp\left(\int\limits_0^{\infty}R(x)\,e^{iwx}\,\mathrm{d}x\right.\right.
 \\\left.\left.- \int\limits_{0}^{\infty}R(x)\,\mathrm{d}x
 + i\mu x - {\sigma_{\rm n}^2\over 2}w^2\right)\right\}(d),
\label{eq:pdfwn}
\end{split}
\end{equation}
where $\mu = -\int x R(x)dx$ for a zero-mean distribution of pixel
values in the map. In general $\mu$ is not known and so is a free
parameter to estimate (or marginalize over). Note that we do not
require the instrumental noise to be white. The model prediction for
the probability distribution is valid even in the case of correlated noise,
provided it is described by Gaussian statistics. In that case the variance
$\sigma_{\rm n}^2$ in the above equation would then be given by the
integral of the noise power spectrum.

\begin{figure*}[!t]
  \begin{center}
    \includegraphics[width=0.4\linewidth]{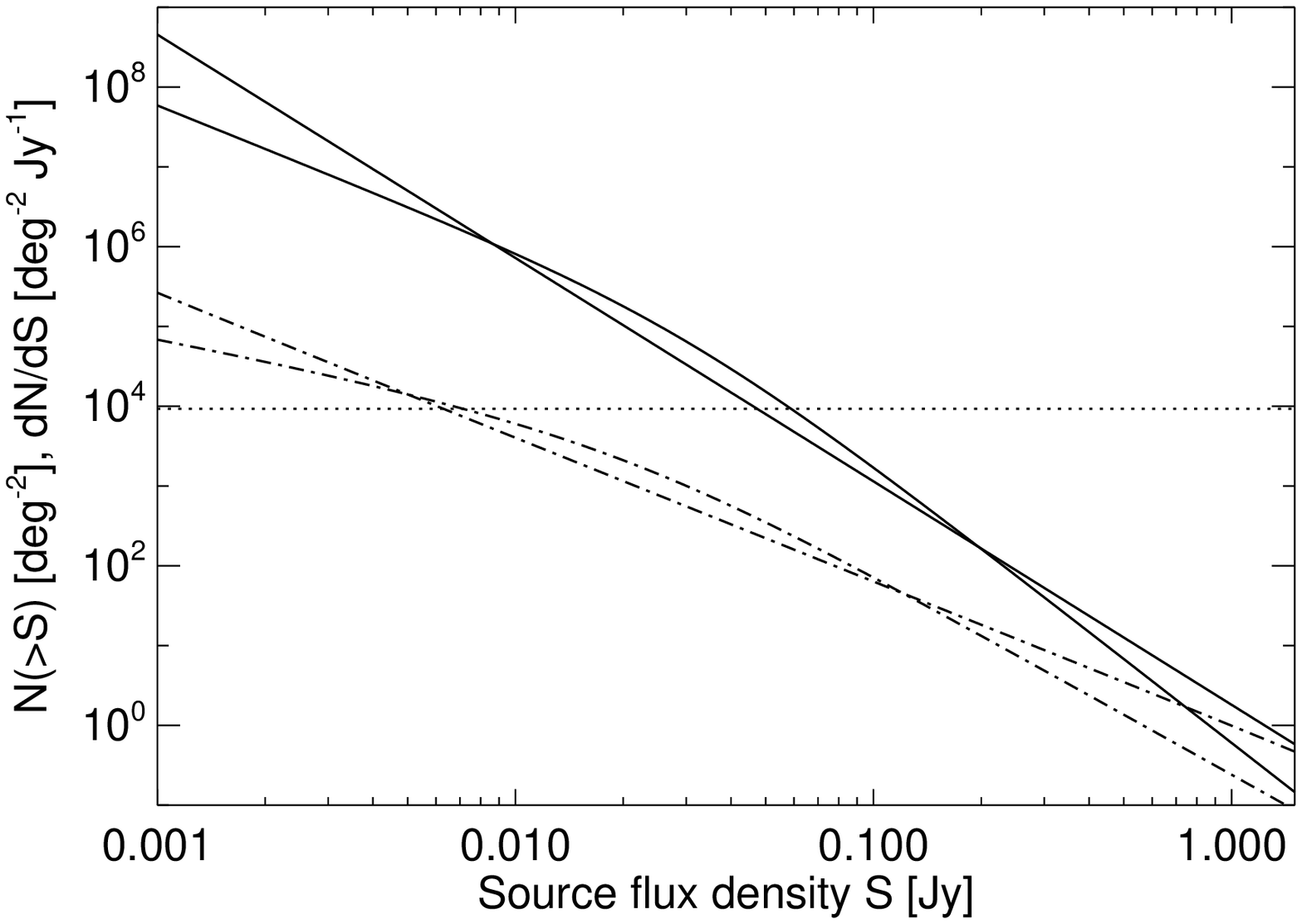}
    \includegraphics[width=0.4\linewidth]{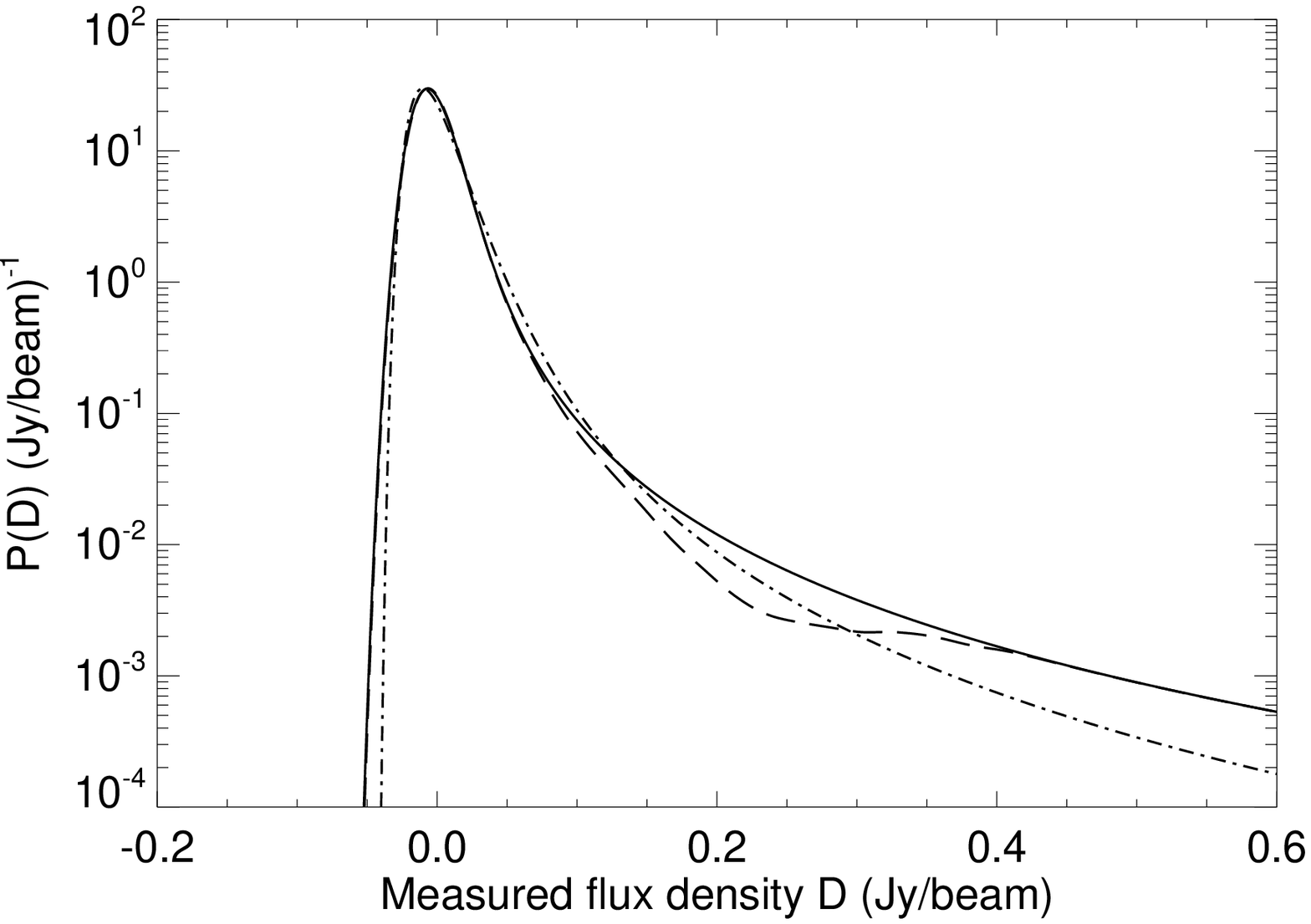}
  \end{center}
  \caption{The left panel shows two examples of differential number
    count models at 250$\,\mu$m (solid curves) and the corresponding
    integrated number counts $N({>}S)$ (dot-dashed curves). The first
    model is a simple power-law of the counts ${dN/ dS} = N_0
    S^{-2.8}$, and the second model is a double power-law ${dN/
      dS}\propto[({S/S_0})^{1.8}+({S/S_0})^{3.5}]^{-1}$ with the
    transition flux $S_0 = 30\,$mJy. The dotted line corresponds to
    one source per 250$\,\mu$m BLAST beam area. The right panel shows
    for the different models the PDFs of pixel values for noiseless
    observations with the 250$\,\mu$m BLAST beam. The solid line
    corresponds to the simple power-law model, the dot-dashed line to
    the double power-law model, and the dashed line to the same single
    power-law model except that for sources between 0.2 and 0.4\,Jy
    the differential counts have been set to zero (this is of course a
    very unrealistic model but the curve is informative nonetheless).
    One can see that at high fluxes the PDFs behave like the
    differential counts power-laws (see text).  Note also that a very
    sharp transition in the counts appears a lot smoother in the PDF,
    even at relatively high flux densities.  Even if no source with
    $0.3\,$Jy flux density is present, a significant number of
    $0.3\,{\rm Jy}\,{\rm beam}^{-1}$ pixels is expected, due to the
    probability of landing near bright sources, as well as the
    probability of having smaller flux galaxies on top of each other.
    This also explains the correlations in the estimated number counts
    parameters described in \S~\ref{sub:degeneracies}.}
 \label{fig:PDFs}
\end{figure*}

Figure~\ref{fig:PDFs} shows the predicted PDF of noiseless
observations with the 250$\,\mu$m BLAST beam for two different galaxy
number count models: a typical 2 power-law model at 250$\,\mu$m
\citep{Borys03} and a single power-law model.
The curves illustrate how sources with different fluxes
contribute to the PDF.  There are 3 features to note:
\begin{enumerate}
\item The very faint sources, for which the average number is much
  larger than one source per beam, induce an almost Gaussian behavior,
  as the Poisson distribution tends to Gaussian for large numbers.
\item Sources with fluxes for which there is about one source per beam
  have the greatest impact on the profile of the distribution (the
  width in particular). They contribute significantly to much higher
  fluxes in the histogram than their own flux because a large fraction
  of them are superimposed on each other.
\item The bright sources contribute mainly to the positive tail of the
  distribution, but also to the full range of pixel brightnesses,
  since sources can contribute everywhere in the beam.  At the bright
  end, the probability distribution varies almost proportionally to
  the source counts for power-law type counts. This can be easily
  checked from equation~(\ref{eq:pdf}) -- if only bright sources are
  present then $~\exp[\int R(x)~e^{iwx}dx]\simeq 1+\int
  R(x)~e^{iwx}dx$, and so $p(d) \propto R(d)$ which is also $\propto
  n(d)$ ($d \neq 0$) for a power-law (see below).
\end{enumerate}

It is  interesting to consider the case of a single power-law
number counts model consisting of
\begin{equation}
  n(S) = K S^{-\alpha}.
\end{equation}
Here the number of sources per `observed' flux $x$
(equation~(\ref{eq:obsflux})) is also a power-law:
\begin{equation}
  R(x) = \Omega_{\rm b} K x^{-\alpha},
\label{eq:omegabeam}
\end{equation}
where $\Omega_{\rm b}$ is the effective beam solid angle, defined as
\begin{equation}
  \Omega_{\rm b} = \int f(\bold{r})^{\alpha-1} d\bold{r}.
\label{eq:omegab}
\end{equation}
One can see that the Fourier transform of the probability distribution
is analytic in this case. The formalism is described in detail in
\citealt{Condon74}.  The resulting distribution is an alpha-stable
function \citep[see][]{Herranz04}.  This expression may be a good
approximation when the number counts only mildly deviate from a
power-law. However, in general, we need an approach that works for a
much wider range of counts models.

\subsection{Dealing with map filtering}
\label{sub:filtpart}

The effect of a spatial filter being applied to the maps can be
modeled entirely by introducing an effective beam which results from
the convolution of the actual beam with the filter kernel.  In most
applications, a high-pass filter is applied to the maps in order to
suppress any residual large scale noise, and this has been done for
BLAST, see \S~\ref{sec:BLASTobs}.  In this case, the kernel is a
`Mexican Hat' shape which suppresses the average value and low spatial
frequencies, and the resulting effective beam takes both positive and
negative values.  Formally, the signal at a given pixel in the
filtered map may be decomposed as the difference of measurements from
two distinct regions in the sky: $d = d_{+}-d_{-}$, with $d_{+}$ being
the result of the integral from the positive part of the effective
beam only and $d_{-}$ from the negative part only (for which the
absolute value is taken).

The probability distribution of the measurements (pixels) in
the map is then the result of the convolution between the two
individual probability distributions:
\begin{equation}
  p(d) = p(d_{+}) * p(-d_{-}),
\end{equation}
which gives, using equation~(\ref{eq:pdfwn}):
\begin{equation}
\begin{split}
  p(d) = {\cal{F}}^{-1}\left\{\exp\left(
 \int R_{+}(x)~e^{iwx}dx \right.\right.\\\left.\left.
 +\int R_{-}(x)~e^{-iwx}dx - \int (R_{+}(x)+R_{-}(x))dx \right.\right.
 \\\left.\left.+ i\mu x - {\sigma_{\rm n}^2\over 2}w^2\right)\right\}(d).
\end{split}
\end{equation}
Here $R_{+}$ and $R_{-}$ are computed from relation (\ref{eq:obsflux})
for the absolute values of the positive and negative parts of the
beam, respectively.  Consequently, the probability distribution
contains a negative tail, which can be understood by realising that
after high-pass filtering very bright sources induce negative shadows
around their locations in the map.  However, this effect is barely
seen in BLAST histograms, since the high-pass filter applied to the
maps is relatively weak.

\begin{figure}[!t]
  \includegraphics[width=\columnwidth]{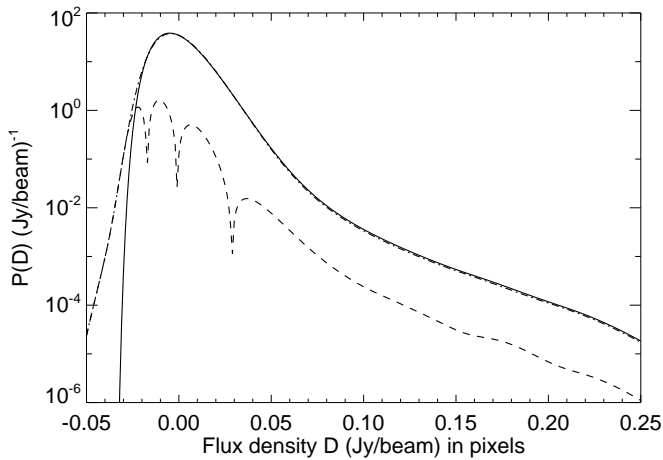}
  \caption{The solid curve represents the predicted PDF (integral
    normalized to unity) of noiseless observations for BLAST at
    500$\,\mu$m derived from our best fit model described in
    \ref{sub:results} and using our best estimate of the beam. The
    dot-dashed curve represents the predicted PDF taking into account
    the effect of high-pass filtering the map, as done in the
    analysis.  The two curves are nearly identical for $x\geq 0$.  The
    dashed line represents the absolute value of the difference
    between the two PDFs which is about 5\% for positive
    $x$. Filtering induces a small negative tail to the histogram and
    affects the high flux part.}
 \label{fig:PDFFilt}
\end{figure}

Figure~\ref{fig:PDFFilt} shows a predicted PDF for noiseless
observations with BLAST at 500$\,\mu$m, with and without high-pass
filtering applied to the maps, as we have done in the analysis (see
\S~\ref{sec:BLASTobs}; the filtering applied is weaker for the other
two wavelengths). 
The effect of filtering is fully taken into account in the analysis of all
BLAST maps presented in this paper.

In developing the techniques described in this section we have in mind
experiments like BLAST which are `total power' (even although the `DC
level' in a map is unmeasurable).  However, the same formalism can be
used to deal with maps resulting from experiments performing
differential measurements, using a chop for example.  In that case the
effective beam can be described as a negative side shifted by some
angular distance from an identical positive side.  The histogram for a
map made with a single difference `double-beam' pattern will be
(statistically) symmetric between positive and negative fluxes, and in
practice this may lead to further complications in inverting the PDF.

\subsection{Number counts parametrization}
\label{sub:model}

We have previously described how to obtain the PDF of pixel values
from a given galaxy differential number counts function assuming that
galaxy locations are uncorrelated. For the analysis of BLAST maps
(described in the next section) we need to do the opposite, i.e., the
objective is to obtain the best estimate of the counts starting from
the determination of the histogram of pixel values, which is a measure
of the PDF. Because the relationship between the counts at a given
flux and the PDF at the same flux is far from straightforward (this is
illustrated in Figure~\ref{fig:PDFs} which shows how the PDF changes
after removing the contribution of galaxies whose fluxes are within a
given flux range), degeneracies are expected between parameters.  It
is therefore necessary to model the counts with a very limited number
of free parameters, as we would do in a deconvolution problem, and to
consider the correlations between these parameters.

We tried different phenomenological models, like single or double
power-laws with a break.  We finally chose to parametrize the
differential number counts by a set of amplitudes at a few predefined
fluxes, with the intervals between flux nodes interpolated with
power-laws to impose continuity of the counts. The number and location
of the nodes are chosen so that the errors on parameters are neither
too small (which would suggest that more nodes could be added) nor too
large, while making sure that the quality of the overall fit is
satisfactory and does not significantly improve with the addition of
more nodes.  For the BGS maps we found that no more than about 6
amplitude parameters per waveband can reasonably be estimated.

\section{Parameter estimation method}
\label{sec:method}

The parameter estimation method developed in this section is similar
to the approach described by \citet{Friedmann04}. It is based on
minimizing the mismatch between the predicted and measured PDFs, i.e.
histograms of the maps, in order to estimate the number count
parameters and also the noise parameters. The approach is based on
maximizing the likelihood of the data given the model.  We denote by
$\theta$ the parameters of the number counts model, which can take any
form.  Later we will focus on a specific model using a handful of
nodes joined with power-laws.

\subsection{Likelihood of the data}
\label{sub:Likelihood}

Let us assume that the different measurements in different pixels of
the map are independent. We will discuss this assumption in
\S~\ref{sub:ApproxL}. The likelihood of the data is the product of the
probabilities of the individual measurements:
\begin{equation}
  L(d|\theta) = \Pi_k p(d_k|\theta),
\label{eq:likelihood}
\end{equation}
where $d = \{d_1, ... d_k, ...\}$ groups all the measured flux values
in the different pixels, and $p(d_k|\theta)$ represents the PDF of
individual measurements given by equation~(\ref{eq:pdfwn}).  However,
instead of equation~(\ref{eq:likelihood}), it will be easier to
consider the log-likelihood:
\begin{equation}
  \log L(\theta) = \sum_k \log(p(d_k|\theta)).
\label{eq:loglikelihood}
\end{equation}
Assuming that the noise is stationary, and that the probability
distribution does not vary much over a pixel flux density bin, then
equation~(\ref{eq:loglikelihood}) becomes
\begin{equation}
  \log L(\theta) \simeq \sum_i n_i \log(p_i(\theta)) + {\rm{constant}},
\end{equation}
where $n_i$ is the number of pixels with flux densities in the $i$th
flux bin interval, e.g. it is the histogram of the data, and $p_i$ is
the result of the integral of the PDF in the $i$th bin, normalized
such that $\sum_ip_i=1$.  See \S~\ref{sub:results}, where we show how
non-stationarity can be taken into account.  Let us define the
quantity $\Phi$ as the negative of the log-likelihood:
\begin{equation}
  \Phi(\theta) = - \sum_i n_i \log(p_i(\theta)) - \log(N!)+\sum_i\log(n_i!),
\label{eq:multin}
\end{equation}
where $N$ is the total number of measurements (i.e., pixels). Now
$\Phi$ is equivalent to the negative logarithm of a multinomial
distribution function (the last two terms in the equation being
derived from the normalisation of the distribution) and is the
quantity we will minimize in order to estimate the number counts. We
notice the following properties: $\left\langle n_i\right\rangle =
Np_i$; ${\rm Var}\{n_i\}=Np_i(1-p_i)$; ${\rm Cov}\{n_i,n_j\}=-Np_ip_j$
($i\neq j$). If the conditions $n_i \gg 1$ and $p_i \ll 1$ are
satisfied for all bins then the quantity $\Phi$ becomes quadratic in
$n_i$:
\begin{equation}
  \Phi(\theta) \simeq {1 \over 2}\sum_i\left({n_i-Np_i(\theta)\over
  \sqrt{Np_i(\theta)}}\right)^2 + K, 
\label{eq:gausscrit}
\end{equation}
where the normalization constant $K = (N/2)\log(2\pi) + (1/2)
\sum_i\log(Np_i)$. Note this quantity is simply a measure of the
mismatch between the probability distribution function and the
histogram of the data\footnote{\citet{Friedmann04} use a similar
  quadratic statistic (without the normalization term $K$) to fit for
  source count parameters, but we prefer to use the actual
  log-likelihood of the data derived in equation~(\ref{eq:multin}),
  which gives proper weights to the tails of the histogram and allows
  us to define very fine flux density binning.}.

In order to estimate the parameters of the model, we minimize
equation~(\ref{eq:multin}) using a simple Markov Chain Monte Carlo
Metropolis Hastings method (MCMCMH), which allows us to sample the
likelihood around its maximum \citep[e.g.][]{Chib95}.

\subsection{Approximate likelihood and error covariance prediction}
\label{sub:ApproxL}

One of the assumptions made for the derivation of the likelihood in
Eq.~\ref{eq:likelihood} is that observations, i.e., pixel values in the
map, are independent of each other. This is obviously not actually the case,
because the beam correlates the signal in adjacent pixels for well
sampled maps. So does any filtering applied to the maps and residual
large scale noise. However, neglecting this effect does not introduce
a bias in parameter estimation, because correlations in pixels only
correlate measurements of the histogram and do not modify its expected
shape.  Nevertheless, neglecting the beam correlations reduces the
performance of the method; another way to think of this is that
sources will cause many neighboring pixels to be bright, but the
information about the spatial distribution of bright pixels is lost if
one only uses the pixel histogram.  As we will describe below, we deal
with this by smoothing our maps with the beam (see
\S~\ref{sub:dataprep}) and then using an estimate of the effective
number of independent pixels.

Measurements of the curvature of the log-likelihood, derived under the
assumption of the incorrect correlations, will lead to an
underestimate of parameter errors.  This can be corrected, to a first
approximation, by taking for $n_i$ in the negative-log-likelihood
expression (equation~(\ref{eq:multin})) the effective number of
independent measurements in the map in flux bin $i$, which is
approximately the number of pixels in flux bin $i$ divided by the beam
solid angle, measured in pixels\footnote{As described in
  \S~\ref{sub:dataprep}, each map has been filtered by the beam for
  better estimation of parameters.  This means that the effective
  correlation length is the beam width for the noise component and
  slightly more for the signal component (larger by ${\sqrt{2}}$). We
  have chosen to use the beam width as the factor, which should lead
  to a mild underestimation of the errors (by a factor smaller than
  ${\sqrt{2}}$, but this has not been fully quantified with Monte
  Carlo simulations).}.  Also, $N$ should be taken to be the total map
area divided by the beam solid angle. This corresponds to applying to
equation~(\ref{eq:multin}) a factor which is the inverse of the beam
area in pixels. We have applied these corrections for the estimation
of uncertainties in the next section.

The error variance and covariance of parameters are obtained from
sampling of the approximate likelihood with MCMCMH (see
\S~\ref{sub:Likelihood}). We have also made a set of 60 Monte Carlo
simulations of 250$\,\mu$m maps following the full processing
procedure. This has allowed us to check the validity of the error
variance prediction using our simple likelihood correction. We find
that differences between our error estimations and the Monte Carlo
simulations are less that 40\%, but this should be quantified more
accurately with larger sets of Monte Carlo simulations.  Although this
has not been checked explicitly, we expect the likelihood
approximation to also be valid for 350 and 500$\,\mu$m data.  The
likelihood contours between pairs of parameters are also computed from
the output of MCMCMH.

Since we calculate the maximum likelihood estimate of the model, and
assuming that the model holds for some value $\theta_0$ of the
parameters, it is also possible to obtain an estimate of the errors by
computing the second derivatives of the likelihood.  The asymptotic
covariance matrix of the estimates is given by
\begin{equation}\label{eq:crb}
  \mathrm{Cov}(\hat \theta) = \langle
  \{(\hat\theta-\theta_0)(\hat\theta-\theta_0)^{\rm T}\} \rangle \simeq
  J(\theta_0) ^{-1},
\end{equation}
where $\hat\theta$ are the estimated parameters and $J(\theta_0)$ is
the Fisher information matrix, which is, following
equation~(\ref{eq:multin}),
\begin{equation}
  [J(\theta_0)]_{kl} = N \sum_i {\partial p_i \over \partial\theta_k} {\partial p_i\over \partial\theta_l}{1\over p_i}.
\label{eq:Fishertmat}
\end{equation}
In practice, the Fisher matrix is evaluated at the point of
convergence $\hat\theta$. Again, $N$ here is the number of beam
areas in the map.

\subsection{Joint fit of maps with different depths}
\label{sub:jointfit}

In order to derive equation~(\ref{eq:multin}), which gives the
quantity which must be minimized, we have made the assumption that the
probability distribution function is the same for all observations (or
pixels). In practice for BLAST this is not valid because the noise
variance changes across the map. One could always minimize the full
expression in equation~(\ref{eq:loglikelihood}), but that would be
extremely time-consuming, since one evaluation of the PDF would be
required per observation. On the other hand, neglecting the
non-stationarity of the noise would be sub-optimal since all pixels
would have the same weight independently of their noise rms.  Ignoring
unequal weights would also bias the parameter estimates, since
non-stationarity leads to a non-Gaussian histogram of the noise part,
even for Gaussian noise. The solution we have adopted is to divide the
observed maps into a limited number of zones such that in each zone
the noise variance is approximately constant. We then compute
histograms and the quantity in equation~(\ref{eq:multin}) for each of
the zones. The resulting criterion to minimize is then
\begin{equation}
  \Phi = \sum_q \phi_q,
\end{equation}
with $\phi_q$ computed from equation~(\ref{eq:multin}) for the noise
variance in zone $q$.

\subsection{Goodness of fit}

The quality of a particular parameter fit can be measured using
Eq.~(\ref{eq:gausscrit}) after rebinning the data such that $n_i \gg 1$
while satisfying $p_i \ll 1$. If the model holds we have the following
relations:
\begin{equation}
  \langle \Phi \rangle \simeq {1 \over 2} (N-n_{\theta}) + K,
\end{equation}
where $n_{\theta}$ is the total number of parameters we estimate, and
$K$ is the normalization term in (\ref{eq:gausscrit}); and
\begin{equation}
  {\rm{Var}}( \Phi) \simeq (N-n_{\theta}).
\end{equation}
These relations can be used to test the compatibility of any given
model with the data.

\section{Application to BLAST number counts}
\label{sec:Appli}

We now discuss the estimation of BLAST number counts using the method
described in the previous section.  First, we discuss the data and
specific processing steps carried out on the maps prior to performing
the likelihood analysis.

\subsection{Data preparation for P(D) analysis}
\label{sub:dataprep}

For source extraction from noisy data it is well known that one should
use a matched filter on a well-sampled map.  It has been shown that
extraction of submillimeter galaxies at low S/N levels can be
performed efficiently by using one of several variants of this
technique -- either thresholding on a beam-correlated map
\citep[e.g.][]{Eales99,Borys03}, or in the case of variable
background, by using a Wiener filter \citep[e.g.][]{Perera08} or its
approximation via a Mexican-hat \citep[e.g.][]{Barnard04,Chapin08}.
This works because beam-fitting is mathematically identical to finding
maxima in beam-correlated maps, provided that the noise is white and
the sources are unresolved.  The Wiener filter simply suppresses any
additional large-scale correlated noise, corresponding to a negative
ring in the spatial filter.

For $P(D)$ analysis, the problem is very similar.  In the low S/N
regime (as in BGS-Wide), the histogram of pixel values will be
dominated by noise, providing that pixels are small enough to fully
sample the beam, which corresponds to 10$\arcsec$ pixels for BLAST.
The histogram depends on the chosen pixel size, which is obviously not
satisfactory because then one is led to either choose between or
combine results obtained using different pixel sizes in order to learn
about structure in the map.  Essentially, in a noisy fully sampled map
there is information about faint structure which is encoded as a
correlation between adjacent pixel values and this information is
ignored in $P(D)$ analysis.  However, after filtering a well-sampled
map with the beam kernel, the noise contribution to the width of the
histogram is reduced by a factor which is approximately the number of
pixels per FWHM of the beam\footnote{Recall that the area of a
  Gaussian beam $2\pi \sigma^2\simeq1.14\, {\rm FWHM}^2$.}, whereas
the signal contribution is only reduced by approximately $\sqrt{2}$,
at least for a Gaussian beam.  It then seems clear that in the
noise-dominated regime, it is better to cross-correlate the maps with
the beam kernel before $P(D)$ analysis.  Even though the map itself is
likely more confused after this process, the resulting histogram
effectively contains information about sources contributing over
roughly a beam area of neighboring pixels.  We have verified that this
improvement is realised in practice, the errors on number count
parameters being reduced by a factor of about three at 250$\,\mu$m for
the bright end fluxes, and by a larger factor at longer wavelengths
and/or for fainter fluxes (after beam convolution of 10\arcsec pixel
maps).

In the opposite regime of source confusion dominating over detector
and sampling noise, it may be a better approach to partially {\em
  deconvolve\/} the map.  In the extreme limit of a noise free map,
one would simply de-convolve to obtain $\delta$-functions and count
the point source strengths. Probably the deconvolution criterion
should be that confusion noise becomes of order the instrumental
noise. However we have not carried out the extensive simulations which
would be required to test this hypothesis.  A relatively good choice
might be to apply a Wiener filter to the confused the
map. Nevertheless, there would appear to be no generally applicable
solution, since one filter might be optimal for the low flux sources,
whereas some different filter might be better for high flux sources.
In practice these filters will depend on the counts themselves
(perhaps related to $\Omega_{\rm b}$ from equation~(\ref{eq:omegab},
for example), so ultimately one might be forced to iterate to find the
best solution.

We have chosen to filter the BGS-Deep+Wide maps simply with the beam
kernels at the appropriate wavelengths, even if these are not strictly
speaking the optimal filter for each map.  We reiterate that this
choice will not bias our results, it just means that our error bars
are not optimal.

\begin{figure*}[!t]
  \includegraphics[width=0.4\linewidth]{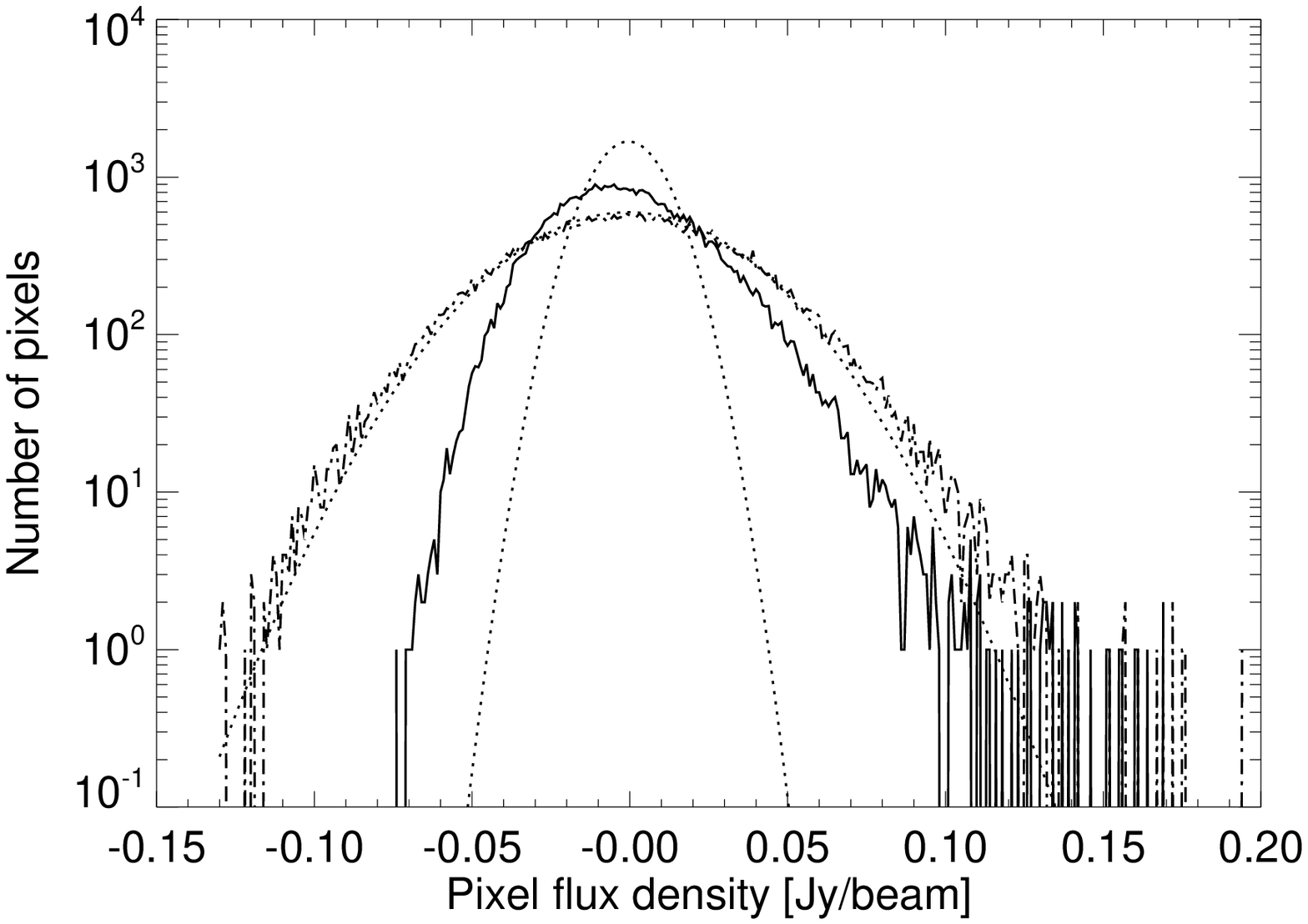}
  \includegraphics[width=0.4\linewidth]{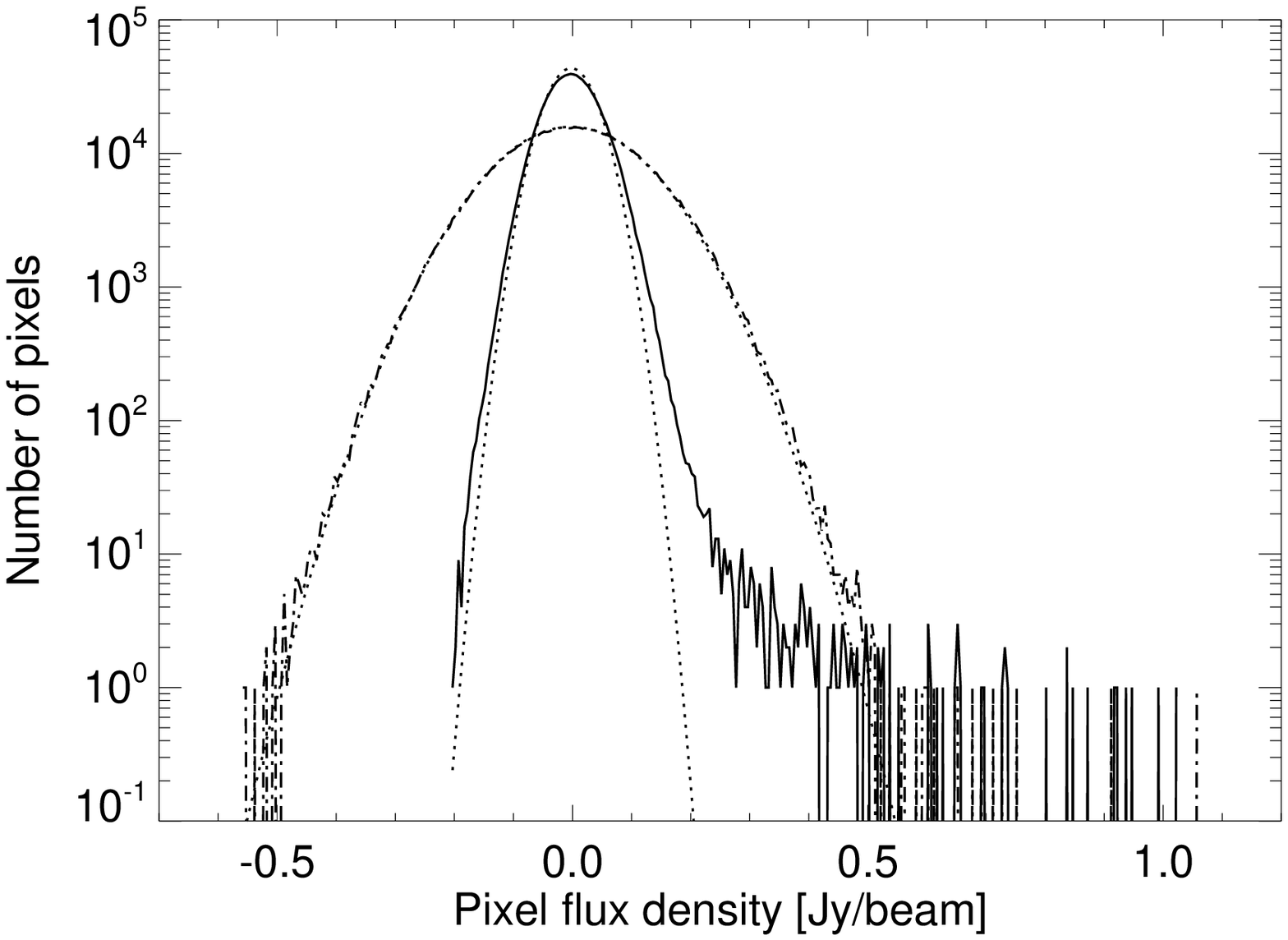}
  \caption{Histograms of the BGS-Deep map (left panel) and BGS-Wide
    map (right panel) at 250$\,\mu$m, before and after
    cross-correlation with the beam kernel (dot-dashed and solid
    curves, respectively). The original map is computed with
    10$\arcsec$ pixels, which is slightly less than one third of the
    pixel size, and the units are Jansky per resolution element in
    each map.  Dotted curves show the expected histograms for noise
    only (both filtered and unfiltered).  The comparison shows that
    filtering the maps with the beam is a much better choice than
    using the raw 10$\arcsec$ pixel map (even in the deep part where
    the signal to noise is larger), in the sense that the difference
    between the measured histogram and the noise variance is enhanced.
    This is because signal from sources is less affected by filtering
    than noise, which has power at higher frequency (see text).}
 \label{fig:filtVSnofilt}
\end{figure*}

Figure~\ref{fig:filtVSnofilt} shows the histogram derived from the
BGS-Deep and Wide map at 250$\,\mu$m using 10$\arcsec$ pixels, before
and after filtering with the beam kernel.
The units of the two maps are Janskys per beam, such that if there was
a source at a given location in the maps, a measure of its flux in
Janskys could be read straight from the pixel at that location. The
expected noise histograms, given the noise variance in each map, are
also shown (also before and after filtering).  From comparing these
curves, we can anticipate a large improvement for $P(D)$ analysis in
using the filtered map with the beam kernel, as compared to using the
raw 10$\arcsec$ pixel map (which is equivalent to filtering the map
with a 10$\arcsec$ square filter), because the width of the histogram
is significantly reduced. The histogram is dominated by noise in the
unfiltered 10$\arcsec$ pixel map, and by confusion in the filtered
map. For BGS-Wide the beam kernel is clearly a very good filter, since
the map is dominated by noise even after filtering.

\begin{figure}[!t]
  \includegraphics[scale=0.58]{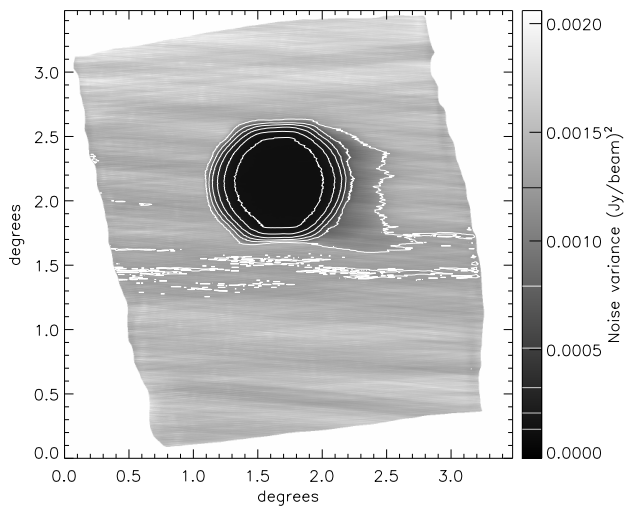}
  \caption{Variance map at 250$\,\mu$m. Contours delimit the different
    variance zones, except the higher variance one which concern a
    small fraction of pixel spread over the top of the map.}
 \label{fig:mapvar}
\end{figure}

Because of the large range of noise variance across the BGS field, we
have divided the maps into 8 zones such that in each zone the noise
variance can be assumed to be constant, following the approach
described in \S~\ref{sub:jointfit}.  The boundaries were chosen so
that variation of the variance in each zone has an rms of about
$\pm$10\%.  Thus, the fluctuations do not significantly modify the
Gaussian shape of the expected noise histograms.  In the end, most of
the constraint on the number count parameters comes from two zones
only: the deepest, containing ${\sim}\,$5\% of the total number of
observed pixels; and a second zone, which covers most of BGS-Wide and
contains ${\sim}\,$80\% of the pixels.  Figure~\ref{fig:mapvar} shows
the variance map at 250$\,\mu$m with our eight regions superimposed.
We can clearly see the two main variance zones.  The other selected
zones are, for the most part, located at the transition between Deep
and Wide.  We could have excluded them from our analysis with little
effect on the parameter constraints, but there is no reason not to
include this additional information, provided these regions are
treated correctly.

The variance in the noise per pixel is computed in the map-making
procedure by propagating the information from estimates of the
timestream noise power spectra (see \citealt{Patanchon08} for
details). Some approximations are made in the calculation, in
particular, we implicitly assume that the residual noise in the final
map is white.  However, we know that map-making is not 100\%
successful in removing large scale noise in these data, partly due to
the low cross-linking angle of the scans. Therefore, even after high
pass filtering the map, a small fraction of correlated noise is
expected to remain.  It is important to notice that residual large
scale noise simply increases the width of the expected noise
histogram, but does not change its Gaussian shape, assuming the input
noise was Gaussian.  In order to minimize potential biases in the
number count parameters which might arise from underestimating the
noise variance in the map, we choose to estimate the noise variance in
every zone except the deepest one as additional free parameters.  We
justify the exception of the deepest zone below. This approach should
be conservative, and in the end it has a very low impact on parameter
estimation errors because the faint end counts, which are mostly
degenerate with noise in the noisier regions, are mainly measured in
the deepest region. Consequently, noise variances in the other zones
are very well determined through the $P(D)$ analysis and fixing those
would not bring significant additional constraints on the counts.  At
the faint end of the counts, since confusion dominates over
measurement noise in the deepest region, errors of 10\% in the noise
variance (which is larger than we expect) would bias the estimated
counts by only a fraction of $1\sigma$. We can confidently fix the
noise variance parameter for the deep part to the predicted value
without risk of biasing count parameters. We have found that the noise
variance measured by $P(D)$ in the wide zone is about 5\% larger than
the variance measured in the pre-processing. Differences are larger in
the zones located at the transition between BGS-Deep and BGS-Wide, but
this is expected considering the potential sources of additional
systematics in regions coincident with scanning accelerations.

We have verified the statistics of noise using jack-knife maps which
were made by computing the difference between two independent maps
from the same region observed at different period during the
flight. We did not find any evidence for departures from Gaussian
instrumental noise when examining the histogram of the difference map.

To build the histograms of data which are used for the fit, we use a
very fine binning in flux density (${<}\,1\,{\rm mJy}\,{\rm beam}^{-1}$)
such that variations
of the PDF within bins is completely negligible. For such small bins
the results are independent of the bin size. As a result, in practice
a large fraction of high flux bins receive either 0 or 1 hit.

\subsection{Estimated number counts}
\label{sub:results}

$P(D)$ analysis is carried out by minimizing the
negative-log-likelihood in equation~(\ref{eq:multin}).  We first
attempt to fit a single power-law model for the number counts: $dN/dS
= N_0 \times (S/S_0)^{\beta}$. Best fit amplitudes and power-law
indices for each of BLAST's three wavelengths are given in
Table~\ref{tab:SinglePLcounts}.

\begin{deluxetable*}{llll}
  \tabletypesize{\footnotesize}
  \tablewidth{0pt}
  \small
  \tablecaption{Best fit power-law number counts \label{tab:SinglePLcounts}}
  \startdata
  \hline
  \hline
\T  \B & {250\,\micron} & {350\,\micron} & {500\,\micron} \\
  \hline
\T  $\beta$      & $-3.005\pm0.022$ & $-3.119\pm0.024$ & $-3.101\pm0.024$\\
  $S_0$ (mJy)   &  7.5 & 2.2 & 1.8 \\
  $\log N_0$ [deg$^{-2}$ Jy$^{-1}$] & 6.036 $\pm$ 0.009 & 7.383 $\pm$ 0.012 & 7.269 $\pm$ 0.015\\
  $S_{\rm min}$ (mJy) & 0.1 & 0.05 & 0.025 \\
  $S_{\rm max}$ (mJy) & 1500 & 550 &  250
  \enddata
  \tablecomments{Fitted models of differential number counts: $dN/dS =
    N_0 \times (S/S_0)^{\beta}$ for the three wavelength. The log of
    the amplitude $N_0$ and the index $\beta$ are both estimated. The
    reference flux density $S_0$ is chosen for each wavelength such
    that errors on the two parameters are nearly decorrelated. We have
    limited the differential counts between $S_{\rm min}$ and $S_{\rm
      max}$, which are given in the last 2 rows.  The high flux limit
    $S_{\rm max}$ is set to be slightly larger than the brightest
    source observed at each wavelength.  The value of $S_{\rm min}$ is
    somewhat arbitrary; we selected a value which is low enough so
    that if we divide it be a factor of 2 the count slope does not
    change significantly.}  \normalsize
\end{deluxetable*}

We find that the best fit single power-laws are very steep, the index
$\beta$ being $-3.0$ at 250$\,\mu$m and $-3.1$ at 350 and
500$\,\mu$m. The strong departure from Euclidean number counts
($\beta=-2.5$) is an indication of strong evolution in the
submillimeter galaxy population. See \citet{Pascale09} and \citet
{Marsden09} for quantitative measurements of these effects made via
stacking.  The steepening of the counts with wavelength suggests a
significant contribution from high redshift galaxies to the confused
signal in the maps.  However, these results are somewhat sensitive to
the faint end flux cut imposed in the model, due to the steep slope of
the counts.

\begin{figure}[!t]
  \includegraphics[width=\columnwidth]{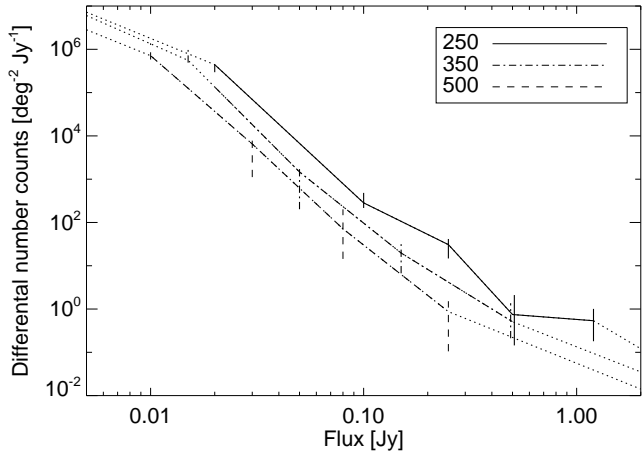}
  \caption{Best fit differential number counts for the three BLAST
    wavelengths. Quoted error bars are derived from the marginal
    distribution for each parameter; because of non-Gaussian behavior
    of the likelihood around its maximum, the best fit model is not
    necessarily centered on the errors bars. The first and last power
    laws (at the faint and bright ends, shown by dotted lines) are
    $3\sigma$ upper limits (corresponding to 99.9\% confidence regions
    for a 1-tailed Gaussian).  There appears to be a shallower slope
    at both ends.}
 \label{fig:countsnopriors}
\end{figure}

The need for a break in the slope towards fainter fluxes, so as not to
overproduce the total cosmic infrared background, shows that the
single power-law model is unrealistic.  Moreover, it appears that such
a simple model is not a very good fit to the data, especially at
250$\,\mu$m, as we will discuss later.  Measurement of a break to a
flat slope at faint fluxes would indicate that the BLAST maps are
nearly deep enough to capture the full intensity of the CIB, and would
presage definitive results from Hershel.  Since one would like to be
able to constrain the number counts over different intervals of source
flux one is led naturally to consider more realistic models.  We have
chosen to fit power-laws for differential number counts within
predefined flux density bins, as described in \S~\ref{sub:model}.  A
number of 6 distinct power-laws are estimated (a total of 7 free
parameters) for the differential number counts at 250$\,\mu$m, and 5
power-laws (6 parameters) at each of 350 and 500$\,\mu$m.  The last
and first power laws are tied to the first and last nodes,
respectively. Number counts are set to zero below the first, and above
the last nodes (they provide limits for the computation of the
PDFs). The choice of flux density for these extreme nodes is set by
requiring them to be very far from typical values constrained by BLAST
(e.g., the extreme nodes are set to $10^{-4}$ Jy and $10$ Jy at
250$\,\mu$m), such that results are independent of our particular
choice.  Best fit number counts for the three wavelengths are
displayed in Figure~\ref{fig:countsnopriors}. The errors bars shown in
the figure are computed from the 68\% confidence intervals on the
marginalized distributions of each parameter separately, the marginal
distributions being estimated by sampling the likelihood with MCMCMH.
Median values of each parameter derived from marginal distributions
and 68\% confidence intervals are given in Table~\ref{tab:counts}.
These are not exactly the parameters of the best fit model, due to
non-Gaussian likelihoods around the maximum.  Pearson correlation
matrices for the parameters are given in
Tables~\ref{tab:corr250}--\ref{tab:corr500}.

\begin{deluxetable*}{ccc|ccc|ccc}
  \tabletypesize{\footnotesize}
  \tablecaption{Best fit differential number counts \label{tab:counts}}
  \startdata
  \hline
  \hline
  \multicolumn{3}{c|}{\T 250\,\micron} & \multicolumn{3}{c|}{350\,\micron} &
  \multicolumn{3}{c}{500\,\micron} \\
 Node  & Best fit & Marginal & Node & Best fit & Marginal & Node & Best fit & Marginal\\
(Jy)  & \multicolumn{2}{c|}{$\log$[deg$^{-2}$\,Jy$^{-1}$]} & (Jy) & \multicolumn{2}{c|}{$\log$[deg$^{-2}$\,Jy$^{-1}$]} & (Jy) & \multicolumn{2}{c}{$\log$[deg$^{-2}$\,Jy$^{-1}$]}\\[1.0ex]
  \hline
\T   $1.0\times 10^{-4}$ & 3.64 & $< 10.28\,(3\sigma)$& $5.0\times 10^{-5}$ & 5.65 & $< 11.12\,(3\sigma)$ & $2.5 \times 10^{-5}$ & 7.47 & $< 11.03\,(3\sigma)$  \\
   0.02 & 5.65& $5.58^{+0.07}_{-0.11}$& 0.015 & 5.75 & $5.88^{+0.14}_{-0.19}$ &0.01 &5.85 & $5.97^{+0.17}_{-0.21}$ \\
   0.1  & 2.45& $2.51^{+0.17}_{-0.18}$&0.05 &3.17   & $2.88^{+0.39}_{-0.57}$ &0.03 &3.81   & $3.63^{+0.36}_{-0.58}$ \\
   0.25 & 1.49& $1.41^{+0.20}_{-0.24}$&0.15 &1.29   & $1.26^{+0.32}_{-0.43}$ &0.08 &1.85   & $1.90^{+0.40}_{-0.75}$ \\
   0.5  &$-$0.13&$-0.10^{+0.43}_{-0.74}$&0.5 &$-$0.29& $-0.23^{+0.36}_{-0.45}$ &0.25 &$-$0.06& $-0.35^{+0.53}_{-0.62}$ \\
  1.2  &$-$0.27&$-0.37^{+0.38}_{-0.36}$& 5 & $-$6.80& $< -2.19\,(3\sigma)$ & 2.5 & $-$12.56 &$< -2.27\,(3\sigma)$\\
  10 & $-$11.77 & $< -3.23\,(3\sigma)$ & & & & & &
   \enddata
   \tablecomments{Differential number counts are parametrized by the
     log of the amplitude at fixed flux density nodes, and filled with
     connected power-laws. Best fit parameters are given for the three
     wavelengths, as well as median values of the marginal probability
     distribution for each parameter. Quoted uncertainties are 68\%
     confidence intervals (except for the first parameters for which
     we give $3\sigma$ upper limits, corresponding to 99.9\%
     confidence), derived from the marginal distribution for each
     parameter. Errors on parameters are highly correlated (see
     Tables~\ref{tab:corr250}--\ref{tab:corr500} for Pearson
     correlation coefficients).}  \normalsize
\end{deluxetable*}

\begin{figure}[!t]
  \includegraphics[width=\columnwidth]{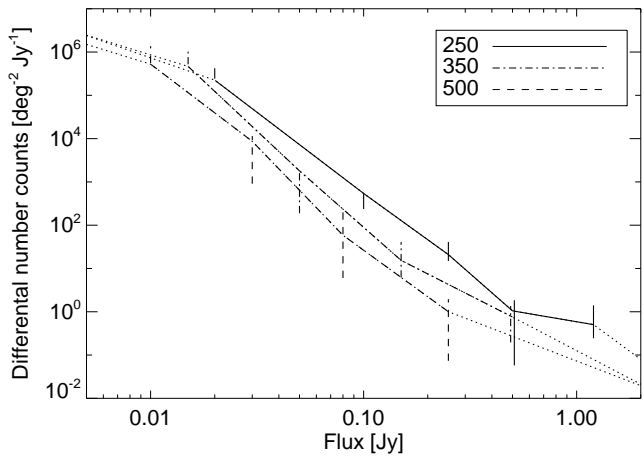}
  \caption{Best fit differential number counts including FIRAS
    background priors. Compare with Figure~\ref{fig:countsnopriors} and see
    \S~\ref{sec:FIRASPriors} for details}.
 \label{fig:countsFIRASpriors}
\end{figure}

\begin{deluxetable*}{ccc|ccc|ccc}
  \tabletypesize{\footnotesize}
  \tablecaption{Best fit differential number counts with background constraint\label{tab:counts+FIRAS}}
  \startdata
  \hline
  \hline
  \multicolumn{3}{c|}{\T 250\,\micron} & \multicolumn{3}{c|}{350\,\micron} &
  \multicolumn{3}{c}{500\,\micron} \\
  Node  & Best fit & Marginal & Node & Best fit & Marginal & Node & Best fit & Marginal\\
\B  (Jy)  & \multicolumn{2}{c|}{$\log$[deg$^{-2}$\,Jy$^{-1}$]} & (Jy) & \multicolumn{2}{c|}{$\log$[deg$^{-2}$\,Jy$^{-1}$]} & (Jy) & \multicolumn{2}{c}{$\log$[deg$^{-2}$\,Jy$^{-1}$]}\\[1.0ex]
  \hline
\T   $1.0\times 10^{-4}$ & 8.44 & $<9.28\,(3\sigma)$  & $5.0\times 10^{-5}$ & 8.32 & $<9.42\,(3\sigma)$ & $2.5 \times 10^{-5}$ & 8.66 & $<9.67\,(3\sigma)$  \\
   0.02 & 5.35 &$5.49^{+0.13}_{-0.11}$ & 0.015 & 5.67    & $5.88^{+0.13}_{-0.17}$ & 0.01 &5.72 & $5.95^{+0.18}_{-0.20}$ \\
   0.1  & 2.73 &$2.57^{+0.16}_{-0.19}$ & 0.05 & 3.26    & $2.85^{+0.39}_{-0.57}$ & 0.03 &3.94 & $3.62^{+0.44}_{-0.66}$ \\
   0.25 & 1.32 &$1.41^{+0.20}_{-0.24}$ & 0.15 & 1.19    & $1.28^{+0.32}_{-0.49}$ & 0.08 &1.77 & $1.87^{+0.52}_{-1.08}$ \\
   0.5  & $-$0.02 &$-0.22^{+0.49}_{-1.01}$ & 0.5 & $-$0.12 & $-0.24^{+0.38}_{-0.46}$ & 0.25 &0.00 & $-0.32^{0.61}_{-0.81}$ \\
\B   1.2  & $-$0.30 &$-0.22^{+0.37}_{-0.39}$ & 5 & $-$10.90 & $< -2.46\,(3\sigma)$ & 2.5 & $-$16.01 & $< -1.75\,(3\sigma)$\\
  10 & $-$21.20 & $< -3.33\,(3\sigma)$ & & & & & &
   \enddata
   \tablecomments{Same as Table~\ref{tab:counts}, but using background
     constraints coming from using the FIRAS measurement as a prior in
     the $P(D)$ analysis. Some of the error bars increase a little
     after adding the FIRAS constraint. This is because some
     parameters are slightly lowered when we add the prior and the
     corresponding amplitudes have a larger probability of being very
     close to zero.}
\end{deluxetable*}

\begin{deluxetable}{c|ccccccc}
  \tablewidth{0pt}
  \tablecaption{Pearson Correlation matrix for the parametrized $dN/dS$ model at 250\,$\mu{\rm{m}}$.  \label{tab:corr250}}
  \tiny
  \startdata
 \B Node & $10^{-4}$ & 0.02    & 0.1     & 0.25    & 0.5      & 1.2 & 10\\ \hline
\B  $10^{-4}$ & 1.00      & $-$0.85 & 0.44    & $-$0.19 & 0.08     & $-$0.06 & $-$0.04\\
\B  0.02      & $-$0.89   & 1.00    & $-$0.77 & 0.42    & $-$0.17 & 0.10 & 0.03\\
\B  0.1       & 0.55      & $-$0.80 & 1.00    & $-$0.67 & 0.28    & $-$0.17& 0.03\\
\B  0.25      & $-$0.25   & 0.43    & $-$0.67 & 1.00    & $-$0.60 & 0.37 &$-$0.07\\
\B  0.5       & 0.04      & $-$0.10 & 0.21    & $-$0.54 & 1.00    & $-$0.72& 0.16\\
 \B 1.2       & $-$0.05   & 0.06    & $-$0.12 & 0.35    & $-$0.73 & 1.00& $-$0.37\\
 \B 10        & 0.02      & $-$0.03 & 0.07    & $-$0.16 & 0.21    & $-$0.39 &1.00 
  \enddata
  \tablecomments{Coefficients are computed for BLAST only (upper
    triangular matrix) and BLAST + FIRAS background constraints (lower
    triangular matrix) following $C_{ij}=\sum_r p_ip_j / {\sqrt{\sum_r
        p_i^2 \sum_r p_j^2}}$, where $p_i$ and $p_j$ are parameter
    number $i$ and $j$, and $r$ is the realization number. Node flux
    units are Jansky.}  \normalsize
\end{deluxetable}

\begin{deluxetable}{c|cccccc}
  \tablewidth{0pt}
  \tablecaption{Pearson Correlation matrix for the parametrized $dN/dS$ model at 350\,$\mu{\rm{m}}$.  \label{tab:corr350}}
  \small
  \startdata
\B  Node [Jy]        &$5\times 10^{-5}$ & 0.015 &  0.05   &  0.15   &  0.5 & 5      \\ \hline
\T  $5\times 10^{-5}$ & 1.00     & $-$0.88 & 0.51     & $-$0.17 & 0.11  & $-$0.04 \\
  0.015             & $-$0.85  & 1.00    & $-$0.75  & 0.33    & $-$0.20 & 0.03\\
  0.05              & 0.43     & $-$0.78 & 1.00     & $-$0.51   & 0.32    &$-$0.03\\
  0.15              & $-$0.16  & 0.39    & $-$0.52  & 1.00    & $-$0.70 & 0.10\\
\B  0.5               & 0.06     & $-$0.23 & 0.35     & $-$0.71 & 1.00 & $-$0.25\\
   5              &  0.02  & 0.02   &   $-$0.05   &  0.20 & $-$0.28    & 1.00
  \enddata
  \tablecomments{Coefficients are computed for BLAST only (upper triangular
    matrix) and BLAST + FIRAS background constraints (lower triangular
    matrix).}
  \normalsize
\end{deluxetable}

\begin{deluxetable}{c|cccccc}
  \tablewidth{0pt}
  \tablecaption{Pearson Correlation matrix for the parametrized $dN/dS$ model at 500\,$\mu{\rm{m}}$.  \label{tab:corr500}}
  \small
  \startdata
\B  Node [Jy]  & $0.000025$ & 0.01 & 0.03    & 0.08    & 0.25 & 2.5 \\ \hline
\T  0.000025 & 1.00     & $-$0.82 & 0.52    & $-$0.34 & 0.14  & $-$0.03 \\
  0.01                & $-$0.71  & 1.00    & $-$0.72 & 0.53    & $-$0.25 & 0.02\\
  0.03                & 0.32     & $-$0.82 & 1.00    & $-$0.69 & 0.39   & $-$0.03\\
  0.08                & $-$0.15  & 0.61    & $-$0.70 & 1.00    & $-$0.59 & 0.03 \\
\B  0.25              & 0.03     & $-$0.33 & 0.49    & 0.63    & 1.00 & $-$0.12\\
  2.5                 & $-$0.01  & $-$0.01 & 0.02    & 0.02    & $-$0.07 & 1.00
  \enddata
  \tablecomments{Coefficients are computed for BLAST only (upper triangular
    matrix) and BLAST + FIRAS background constraints (lower triangular
    matrix).}
  \normalsize
\end{deluxetable}

There appears to be a change towards shallower slopes at the faint
ends of the counts at all three wavelengths, and also at the bright
end at 250\,\micron. This bright end behaviour is consistent with the
expectation that we are entering the Euclidean regime for ${\sim}\,$Jy
sources.  At all 3 wavelengths the counts are much steeper than
Euclidean over most of the range of flux densities probed. The slope
of the counts is close to $-3.7$ at 250$\,\mu$m over the flux density
range range 0.02--0.5$\,$Jy, and $-4.5$ for the two longer BLAST wavelengths in
approximately the same range.  This is not compatible with the slope
values fitted assuming single power laws. The break in the slope
observed at the faint end is consistent with the requirement that the
background not be overproduced.  Since the intensity made by the
counts is $I=\int S.(dN/dS)dS$ then we must have $\beta\,{>}\,-2$ as
$S\to0$ in order to have convergence.  This is not imposed in our
method since the mean value of the probability distribution is set to
0, so we could have found un-physical solutions of $\beta\,{<}\,-2$.
Figure~\ref{fig:countsFIRASpriors} shows that there is some
improvement in the faint end slope constraint, shown with a dotted
line, when we include a prior on the total background intensity.  In
addition, as we discuss in \S~\ref{sec:FIRASPriors}, imposing this
prior also reduces parameter degeneracies.

The gain in the quality of fit going from the single to
multi-power-law model can be evaluated by measuring $\Delta\Phi$,
which is the difference of the minimized quantity for the two
models. Since the single power-law model is contained within the set
of multi-power-law models, we expect $\Delta\Phi/2 \approx 1$ for each
degree of freedom removed. We found that $\Delta\Phi/2$ = 15.5, 8.7,
and 8 at 250, 350, and 500$\,\mu$m, respectively, while the difference
in the number of fit parameters is 5 at 250, and 4 at 350 and 500. The
multi-power-law model is therefore a significantly better
representation of the data. We have also checked that adding more
number count parameters (by dividing the counts into more flux nodes)
does not significantly improve the fit.  We thus conclude that the BGS
data can constrain ${\sim}\,6$ parameters.

\begin{figure*}[!t]
  \includegraphics[scale=0.338]{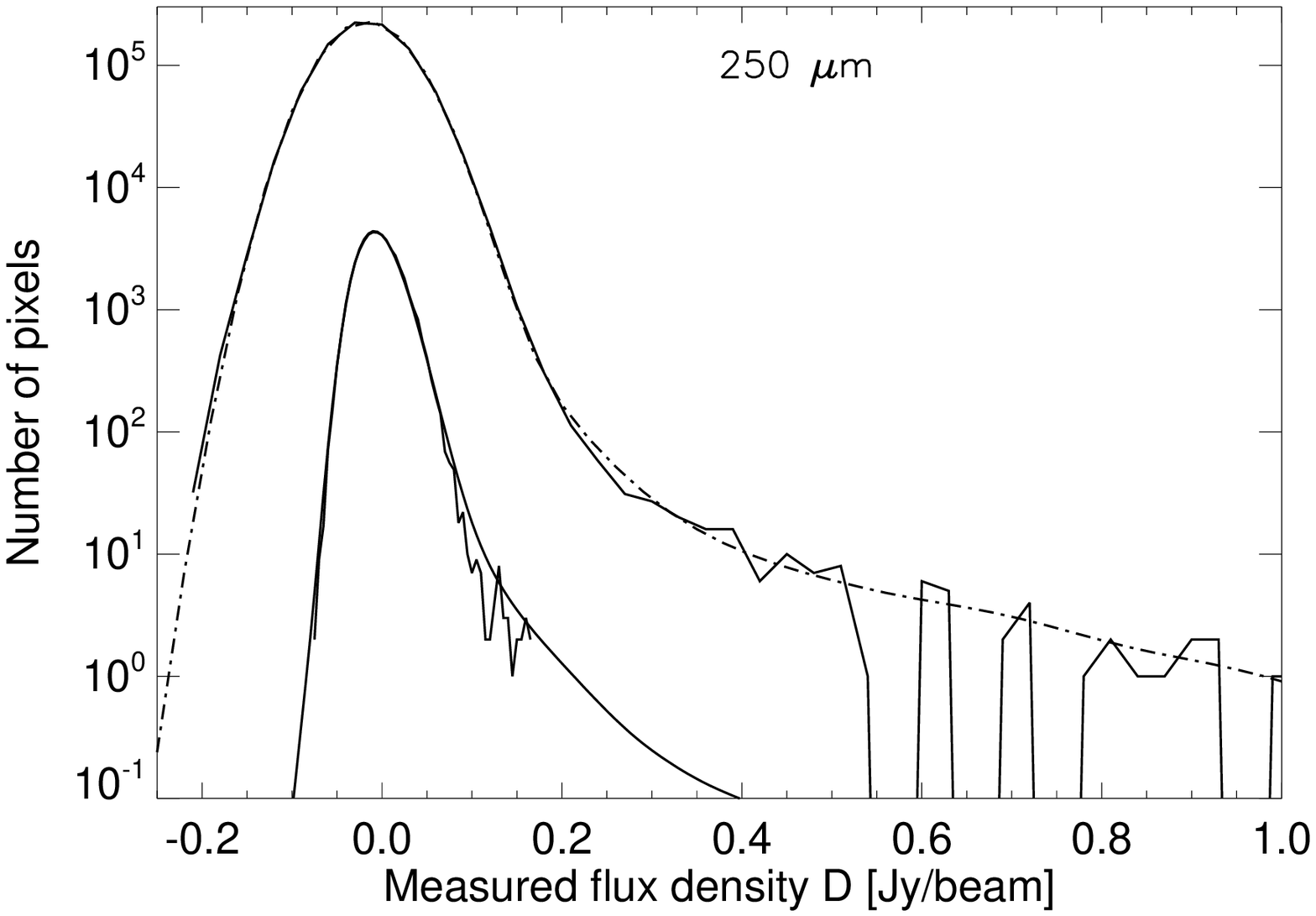}
  \includegraphics[scale=0.338]{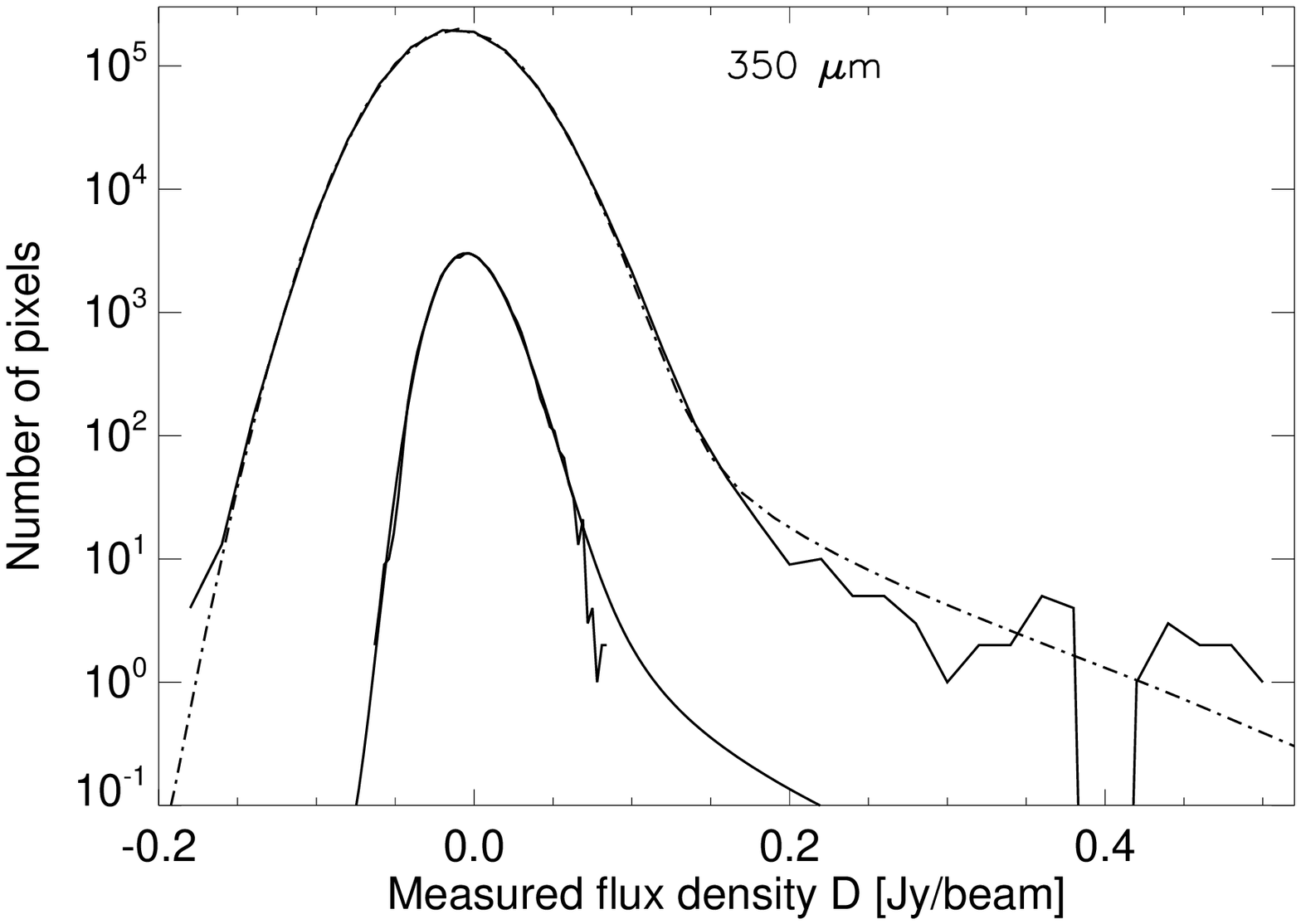}
  \includegraphics[scale=0.338]{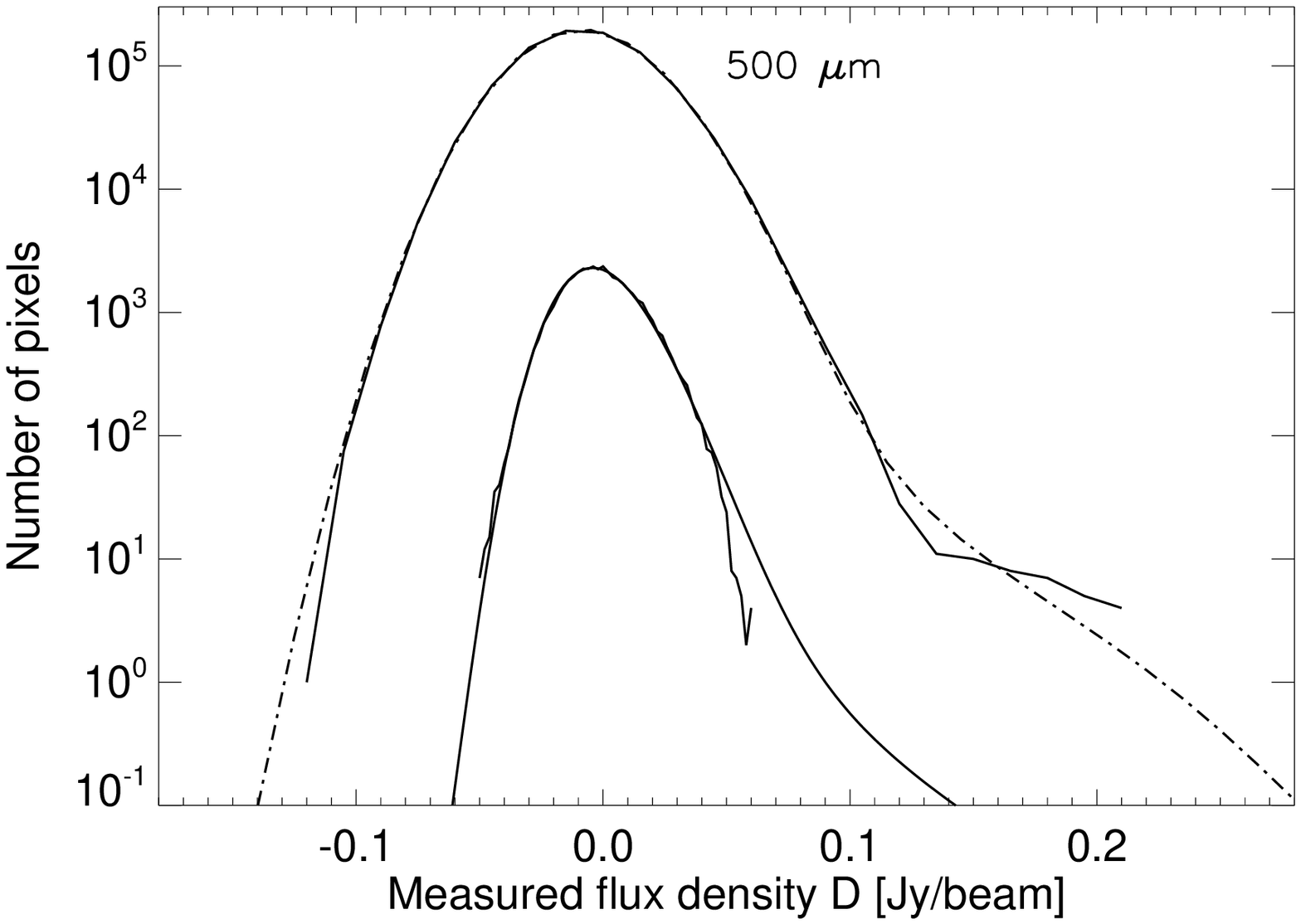}
  \caption{Histogram of pixel values for the deep and wide zones
    (which correspond closely with BGS-Deep and BGS-Wide) compared
    with prediction from the best fit model of the differential counts
    (dot-dashed line for deep and solid line for wide). The bins in
    flux density are chosen to be different for the deep and wide
    histograms for clarity, and are both much larger that the binning
    used for parameter fitting.  This figure shows the very good fit
    of the model to the data at all three wavelengths. The apparent
    discrepancy at the bright end for the histograms of the deep part
    is not very significant, because of the large correlations of
    values between bins in the histograms.  We can see that the deep
    region, which is confusion limited (the contribution of the noise
    to the histogram is smaller than the source confusion
    contribution), contains almost all the information on faint
    sources, which are within the noise regime of the wide histogram.
    At the bright end, the Wide part carries almost all the
    information about sources above 0.2\,Jy at 250\,\micron, and
    0.1\,Jy at 350 and 500\,\micron, showing the importance of sky
    coverage.}
 \label{fig:histodeepwide}
\end{figure*}

We compare the predicted histograms (i.e., rescaled PDFs) of the best
fit multi-power-law model, with the actual histograms of the maps in
Figure~\ref{fig:histodeepwide}.  We separately plot the histograms of
the deep and wide zones and ignore the others here, which give only
weak additional constraints on the parameters.  One can see that there
is a very good match of the model to the data, considering that pixels
of the maps are correlated on the beam scale, i.e., for about 3.5,
4.5, and 6 pixels of $10^{\prime\prime}$ size at 250, 350,
500\,\micron, respectively. The apparent discrepancy in the histograms
of the deep section of the map at fluxes around
$0.1\,{\rm Jy}\,{\rm beam}^{-1}$ at
250\,\micron\ (around $0.05\,{\rm Jy}\,{\rm beam}^{-1}$
at 500\,\micron) is in fact not very
significant, because correlations of pixels in the map induce
correlations in the histogram bins.  However, some of the difference
at the bright end of the deep counts may be real.  On average, a
significant fraction of the pixel flux at around
$0.2\,{\rm Jy}\,{\rm beam}^{-1}$ (say, at
250\,\micron) comes from very bright sources which are observed at the
edge of the beam (250\,\micron\ beam also has significant side lobes,
see \cite{Truch09}; they are taken into account in this analysis), and
it appears that the measured density of such sources in the deep
region is a bit lower than the density observed in the wide
region. This explains why at the bright end, the measured histograms
of the deep sections are systematically lower than the prediction from
the best fit model, even at moderate fluxes.  This is partly explained
through the historical choice of this region (initially the {\em
  Chandra\/} Deep Field South) to be devoid of bright sources
(although at other wavelengths and over a smaller field, making the
strength of any bias hard to assess).

\subsection{Degeneracies}
\label{sub:degeneracies}

We find that the errors on the number count parameters are highly
correlated, and the likelihood around the maximum has a very
non-Gaussian behavior for some parameters.  The correlation is
negative for two adjacent nodes in the counts.  This is expected,
since sources at a given flux contribute to a large range of pixel
fluxes in the map, due to confusion (as illustrated by
Figure~\ref{fig:PDFs} showing how the histogram is modified after
removing the contribution of sources of a given flux), and hence
shifting some sources out of one flux bin and into the next will lead
to two histograms with approximately the same shape. This
anticorrelation is stronger for lower flux density nodes. This is
again completely understood -- faint sources produce a nearly Gaussian
histogram, with no particular structure allowing us to distinguish
between the number of sources and their flux.

The degree of correlation is evaluated using the Pearson correlation
matrix, with results given in
Tables~\ref{tab:corr250}--\ref{tab:corr500}.  However, some precaution
should be taken in using these numbers because of the locally
non-Gaussian likelihood around its maximum. Some error bars are
significantly asymmetric, as seen in Figure~\ref{fig:countsnopriors},
and more elongated on the lower side.  This is due to the fact that
number counts can be zero with a probability which is not entirely
negligible in some flux bins in which case the parameter values, which
are the logarithm of amplitudes, can take very negative values.  We
have set quite low thresholds for the faintest node to prevent
parameters taking values which are too low, based on the physical
assumption that number counts should be described by a relatively
smooth function. The lower errors for some parameters are slightly
dependent on these thresholds. This is a complication introduced by
the choice of parametrizing counts with the log of amplitudes, but
this parametrization seems justified considering the power-law nature
of counts.

\begin{figure}
  \includegraphics[scale=0.2595]{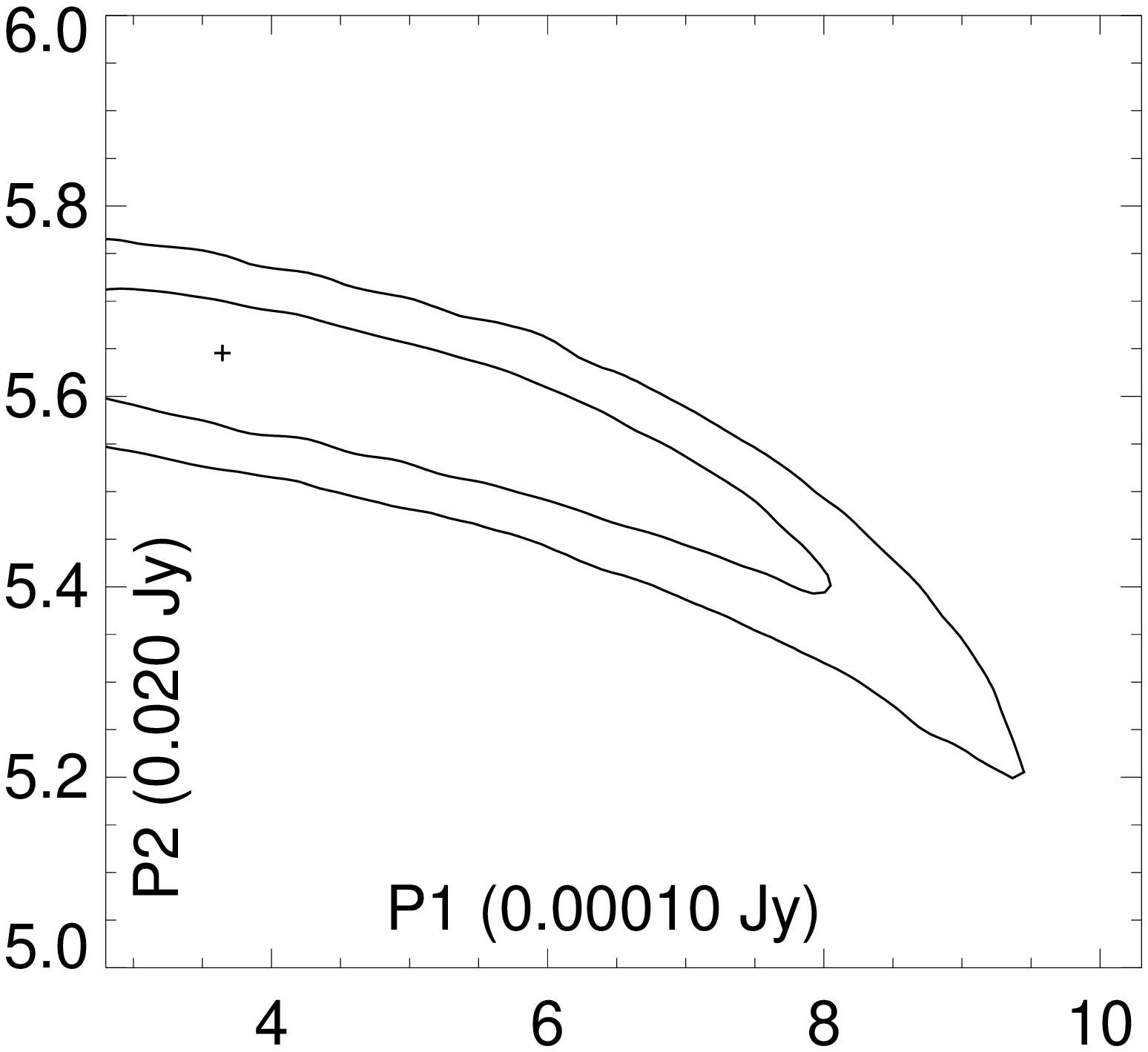}
  \includegraphics[scale=0.2595]{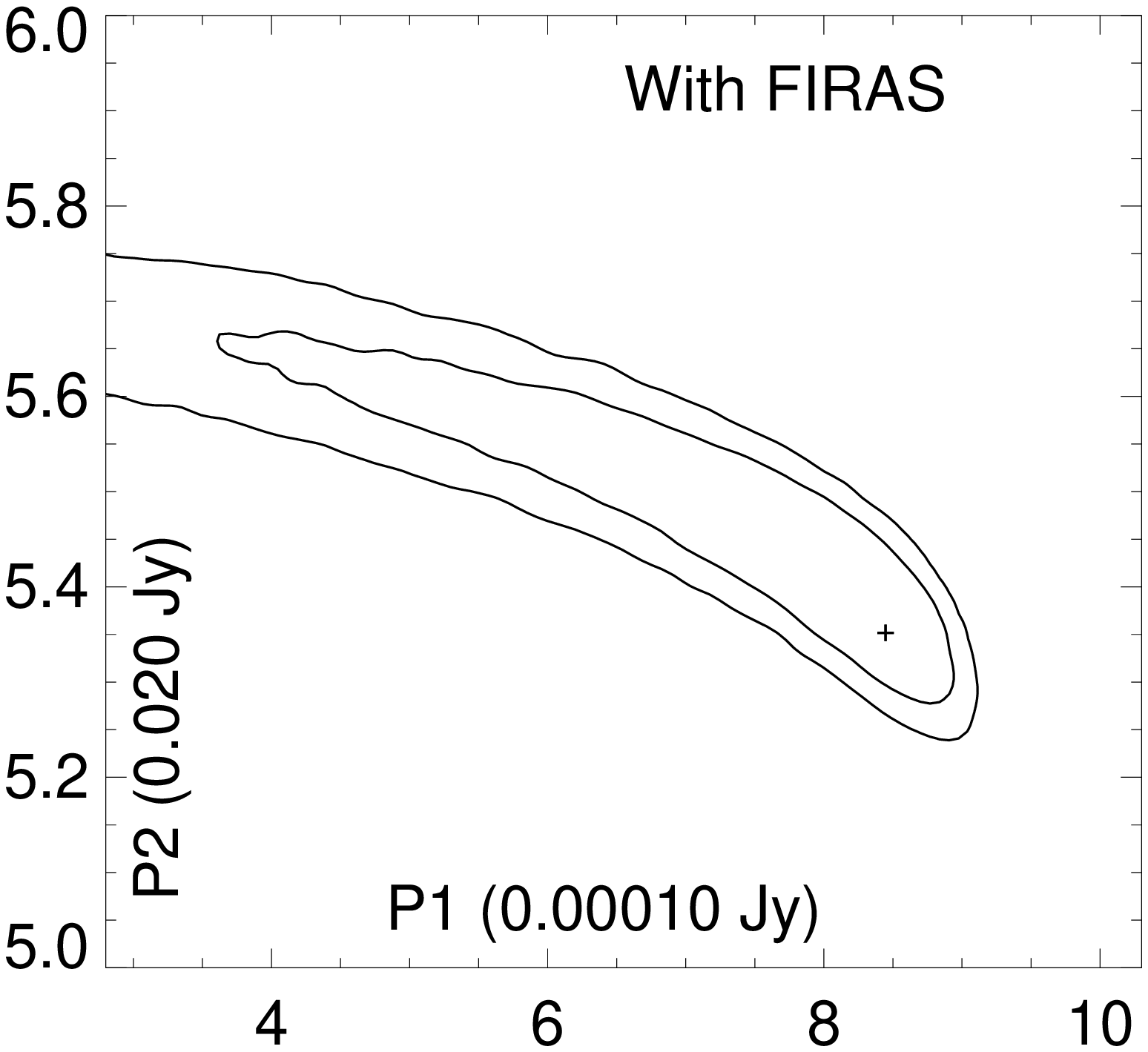}
  \includegraphics[scale=0.2595]{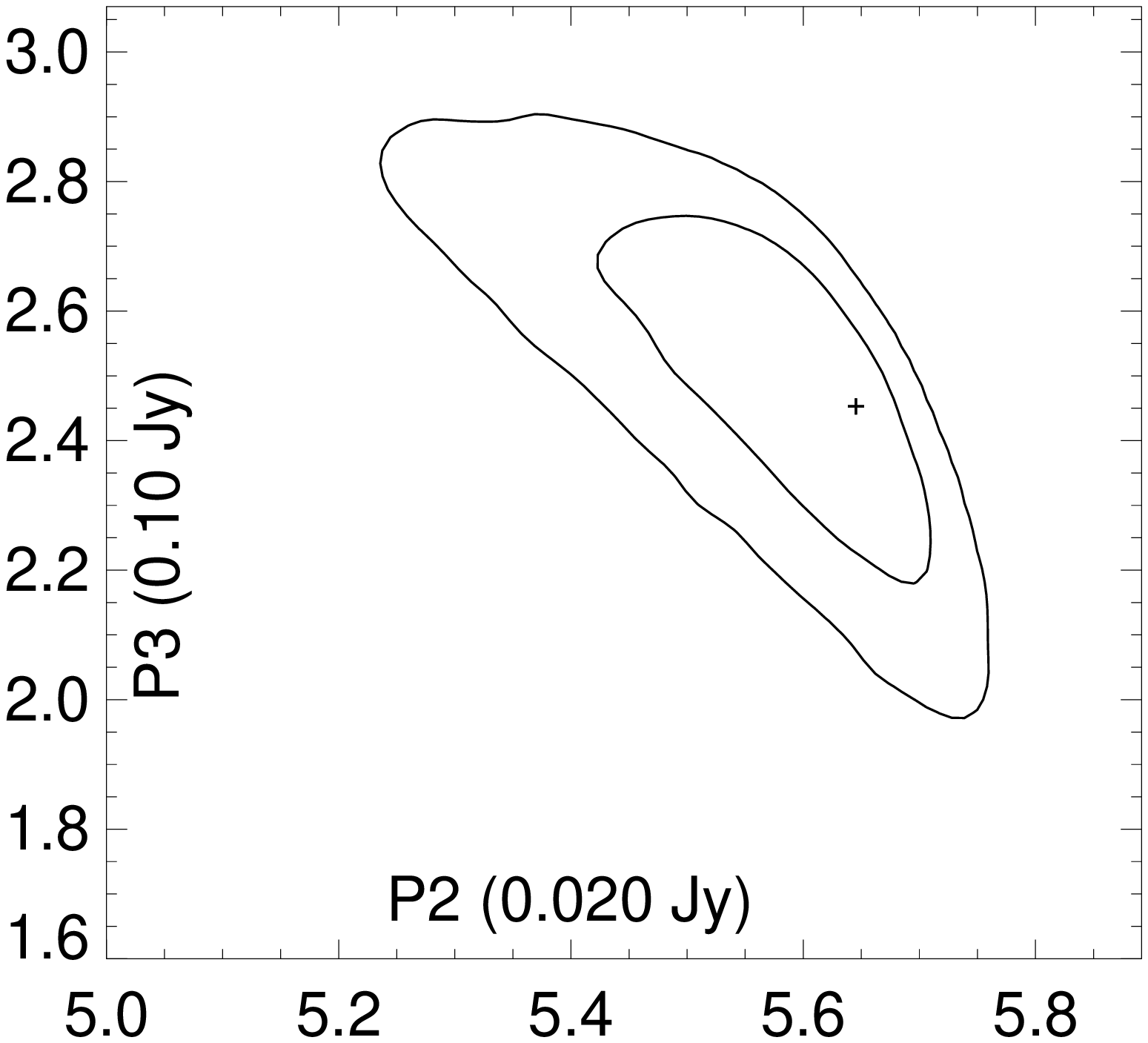}
  \includegraphics[scale=0.2595]{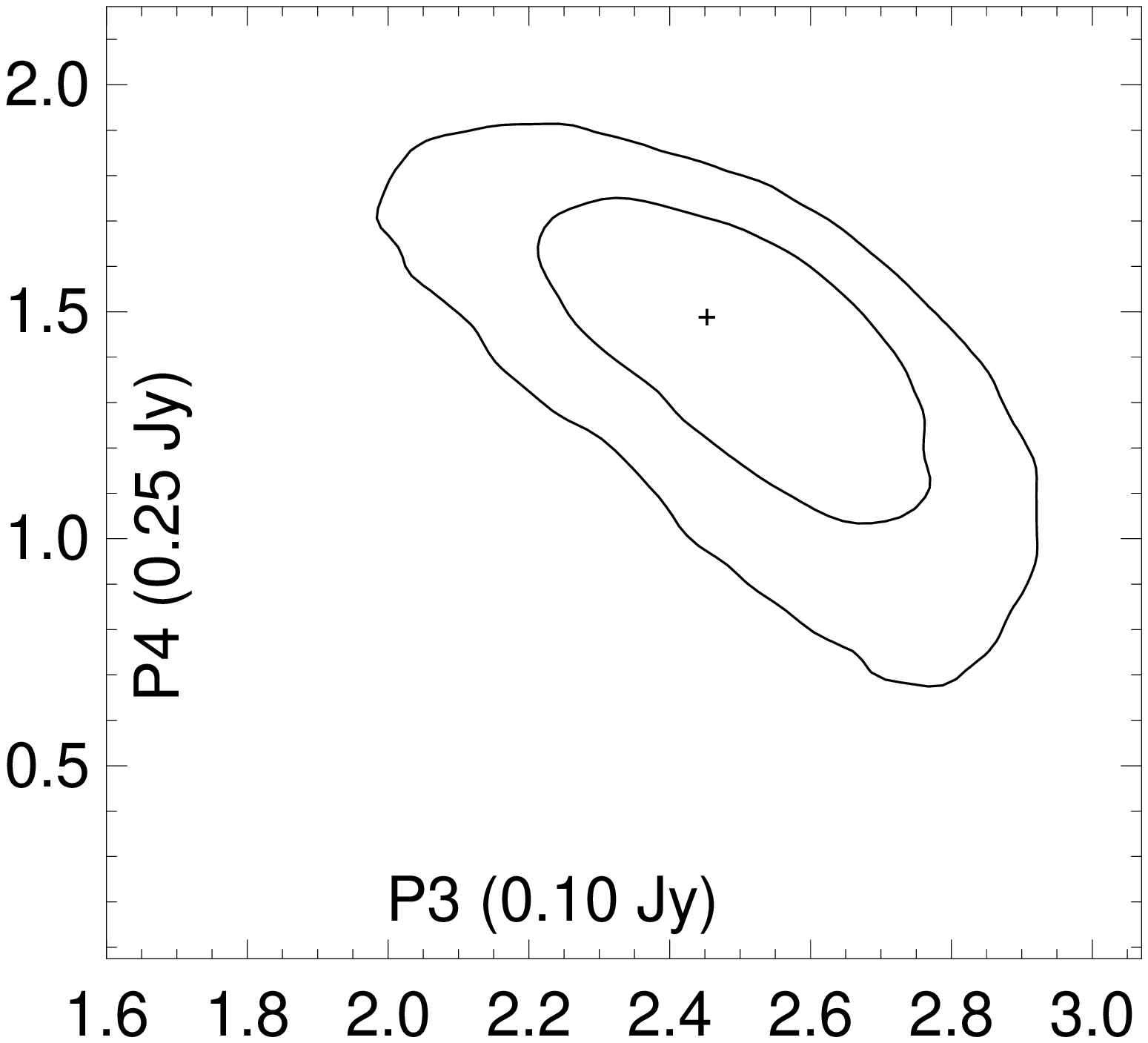}
  \includegraphics[scale=0.2595]{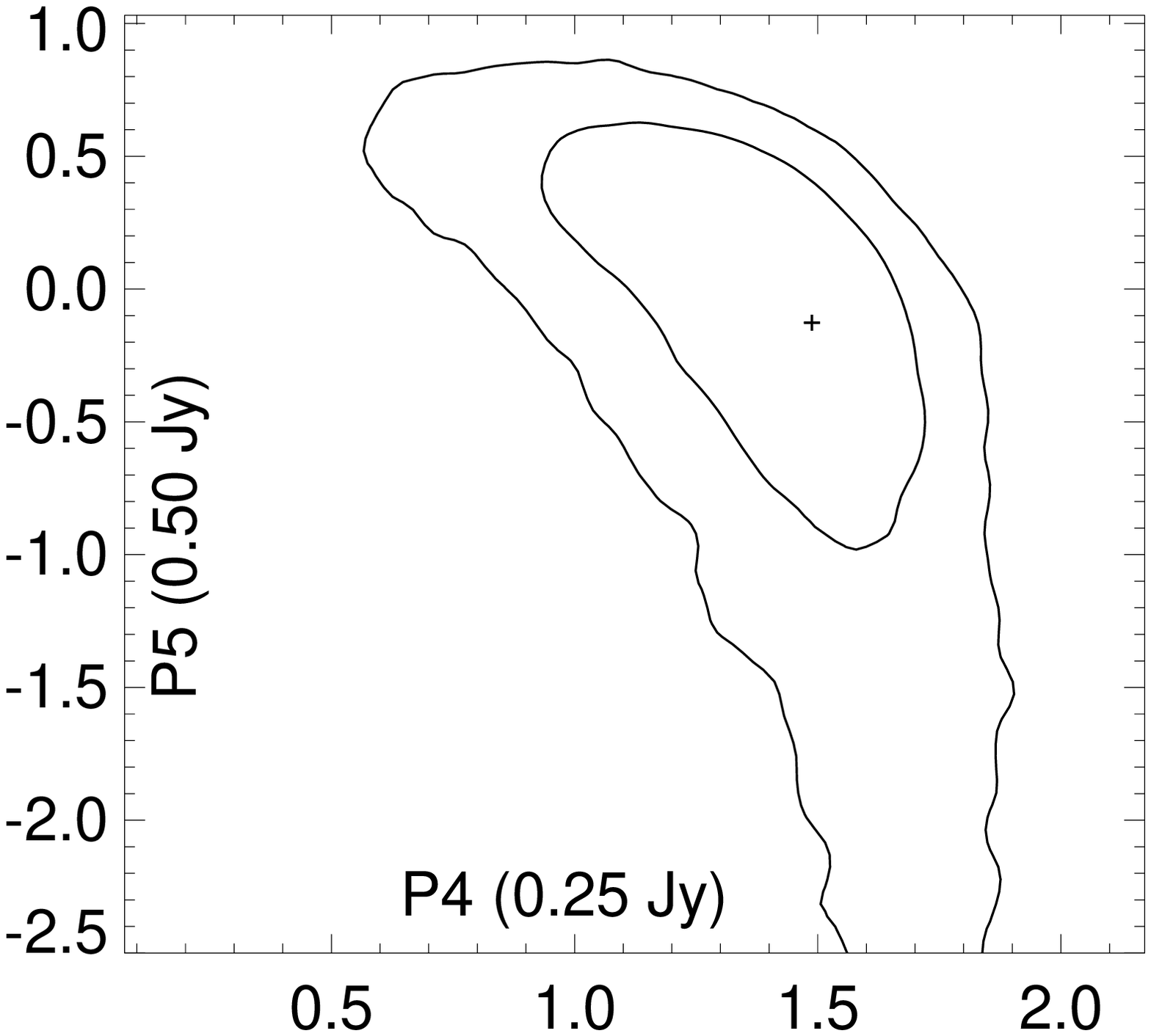}
  \includegraphics[scale=0.2595]{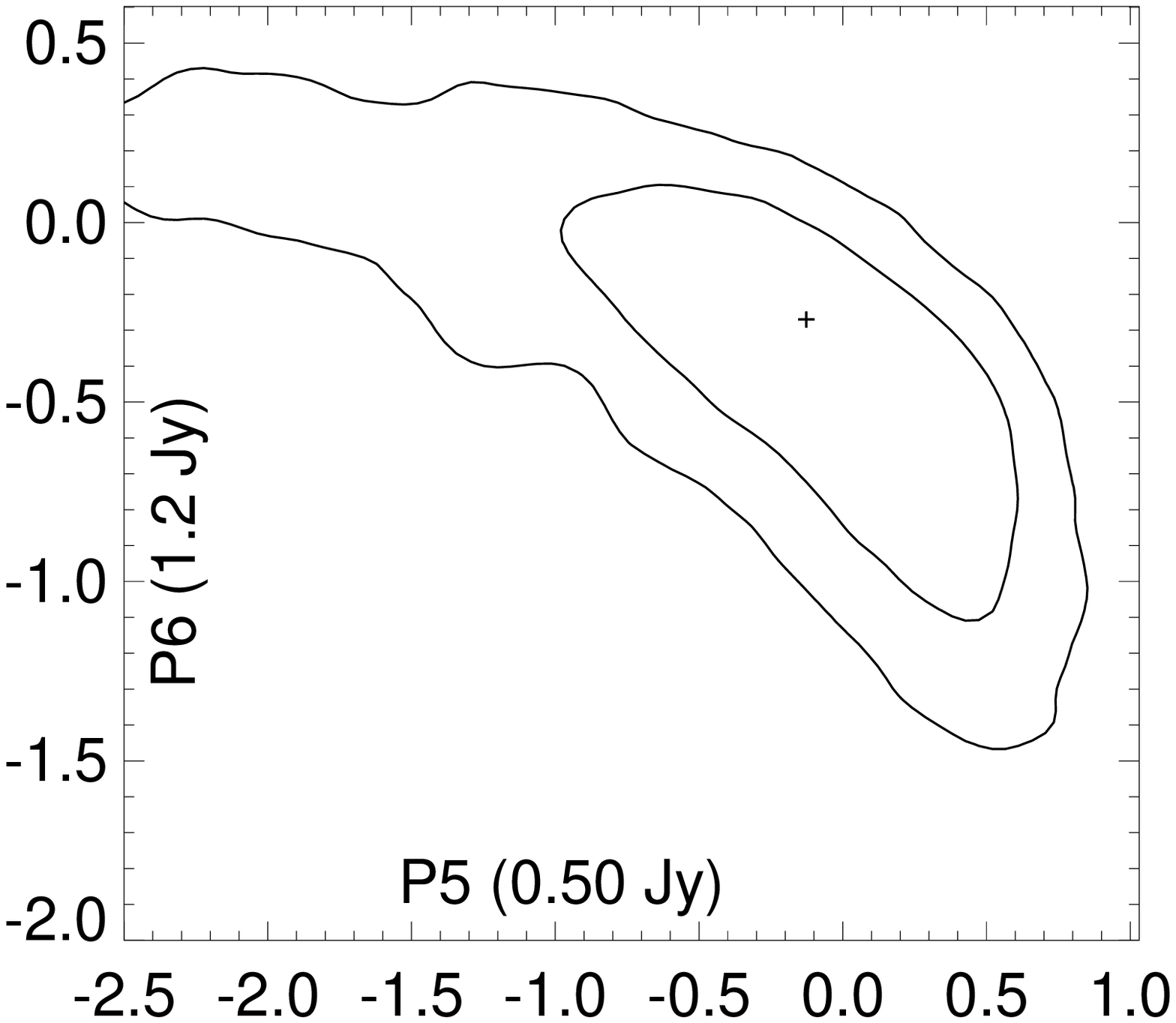}
  \includegraphics[scale=0.2595]{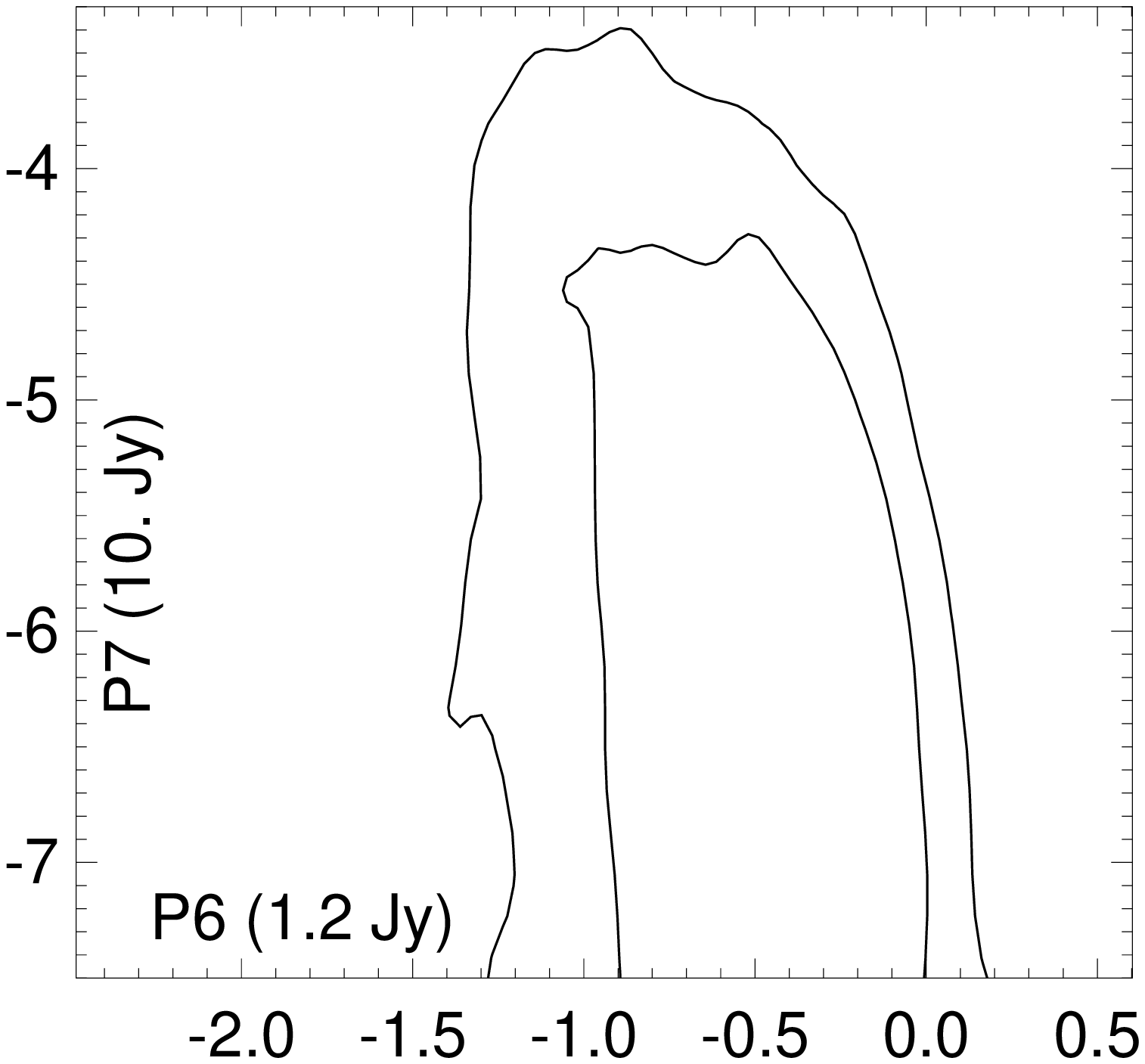}
  \caption{Likelihood contours for pairs of parameters associated with
    adjacent nodes at 250\,\micron. The two curves in each panel
    represent 68\% and 95\% intervals estimated from sampling the
    likelihood with MCMCMH. Parameter units for each graph are
    $\log$[deg$^{-2}$\,Jy$^{-1}$]. All panels are for $P(D)$ analysis
    using BLAST data only, except for the second panel (top right
    side), which shows the likelihood contours for the first two
    parameters (i.e., the faint end nodes) after adding FIRAS
    background constraints. Crosses indicate best fit parameter
    values.  $P(D)$ analysis provides only upper limits for number
    counts at the first and last nodes.  The amplitude of the counts
    at 0.5\,Jy can be zero with a low but not completely negligible
    probability.}
 \label{fig:2DContours}
\end{figure}

Parameter correlations can be seen more explicitly in
Figure~\ref{fig:2DContours}, which shows two-dimensional likelihood
contours for pairs of parameters at $250\,\mu$m.  One can see that
none of the contours are well described by ellipses.

The degeneracies between parameters are such that number counts can be
estimated only in a very limited number of bins (of the order of 6 for
BLAST). The resolution limit of the counts strongly increases with the
resolution of the maps, since improved resolution leads to a more
direct relationship between the probability distribution and the
number counts.  The SPIRE instrument on {\em Herschel\/}, with twice
the resolution of BLAST, should allow significantly finer binning of
the counts and should greatly improve statistics because of
dramatically extended coverage of the sky.

\subsection{Including constraints from FIRAS}
\label{sec:FIRASPriors}

For the three BLAST wavelengths, the lowest flux node can be extremely
low without limit and still give a good fit.  In other words the BLAST
data are compatible with a model predicting no sources in the first
bin. Nevertheless BLAST shows without ambiguity that there is a break
towards the faint end for all three wavelengths. The number of these
fainter sources on the sky has a stronger impact on the total
intensity of the background radiation, which is not directly
measurable by BLAST, than on the width of its distribution. It appears
that some of the models which are compatible with BLAST histograms
produce a CIB which is inconsistent with FIRAS measurements (see
\citealt{Puget96} and \citealt{Fixsen98}) at these wavelengths.

We have used FIRAS constraints published in \citet{Fixsen98} as a
prior in the $P(D)$ analysis in order to select against models which
have an unphysically small abundance of faint sources.  Even though
the FIRAS measurements have quite large uncertainties, the additional
constraint is helpful in breaking this degeneracy.  BLAST number
counts including the FIRAS constraints are shown in
Table~\ref{tab:counts+FIRAS} and in
Figure~\ref{fig:countsFIRASpriors}, with the Pearson correlation
matrix given in the lower triangular parts of
Tables~\ref{tab:corr250}--\ref{tab:corr500}. Estimation of the number
counts is improved at the faint end after adding the FIRAS
constraint. The break in the counts at the faint end is also more
clearly detected here than in the case without priors. As an example
two-dimensional contours including FIRAS priors are shown for the
first two parameters at 250\,\micron\ in Figure~\ref{fig:2DContours}
(top-right panel).

\section{Discussion}
\label{sec:Discuss}

\subsection{Comparison with other data}

At $250\,\mu$m BLAST has provided the only existing images of the sky,
and hence the counts estimates are unique.  However, it is possible
(although challenging) to obtain data from the best ground-based
facilities operating in the $350\,\mu$m and $450\,\mu$m atmospheric
windows.  Several studies have been published using SCUBA;
\cite{Smail02,Borys03,Knudsen06} or SHARC; \cite{Khan07,Coppin08}.
These counts estimates are based on a few to a handful of sources, and
both calibration and reliability are issues for interpreting the data.
By flying above the bulk of the atmosphere, BLAST was able to increase
the number of detections (or overall statistics on the counts) at
these wavelengths by about 2 orders of magnitude.

\begin{figure}[!t]
  \includegraphics[width=0.9\linewidth]{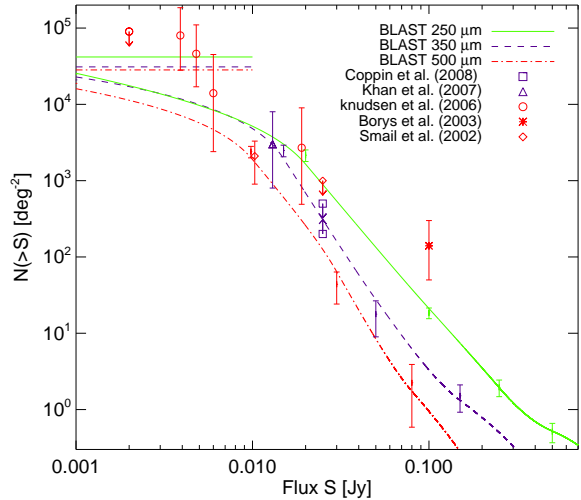}
  \caption{Best fit cumulative (or integral) BLAST counts, compared
    with published estimates using ground-based facilities at
    $350\,\mu$m and $450\,\mu$m. The three curves represent the BLAST
    cumulative counts at the three wavelength (250$\,\micron$ is in
    green 350$\,\micron$ in blue, and 500$\,\micron$ in red) derived
    by integrating the best fit multi-power-law models including FIRAS
    constraints. Interval of 68\% confidence (which appear without
    central dot in Figure) are estimated at each flux node that we
    have defined in our differential count model, by measuring the
    dispersion of the likelihood sampling by MCMCMH. Other data at
    350$\,\micron$ and 500$\,\micron$ are shown in blue and red,
    respectively. Best fit models from BLAST are not necessarily
    centered on error bars due to significant departure of the
    likelihood from Gaussian. The lines indicate upper limits
    of cumulative counts at the fainter nodes of $10^{-4}$\,Jy at
    250$\,\micron$, $5 \times 10^{-5}$\,Jy at 350$\,\micron$ and $2.5
    \times 10^{-5}$\,Jy at 500$\,\micron$. Upper and lower limits from
    other experiments are $1\sigma$ limits. No waveband correction has been
    applied to the $450\,\mu$m counts for the comparison with BLAST
    counts; those measurements should in principle lie somewhere
    between BLAST 350 and $500\,\mu$m counts}
 \label{fig:countsCumulative}
\end{figure}

BLAST data are sufficient to allow us to determine {\em
  differential\/} counts, whereas all published estimates at similar
wavelengths have been cumulative counts, i.e., $N({>}\,S)$.  In
Figure~\ref{fig:countsCumulative} we compare our best fit estimate for
the cumulative counts with the published estimates at $350\,\mu$m and
$450\,\mu$m.  Note that there is an additional calibration uncertainty
when comparing with other experiments (as well as issues of different
waveband filter profiles). BLAST cumulative counts shown in
Figure~\ref{fig:countsCumulative} are computed by integrating the best
fit differential counts, and errorbars are estimated from the
dispersion of count amplitudes at each flux node.

One can see that BLAST provides an accurate measurement of the counts
over a wide range of flux densities. This is due to an unprecedented
signal-to-noise ratio and sky coverage delivered at these wavelengths.

\subsection{Comparison with models}

\begin{figure}[!t]
  \includegraphics[width=0.9\linewidth]{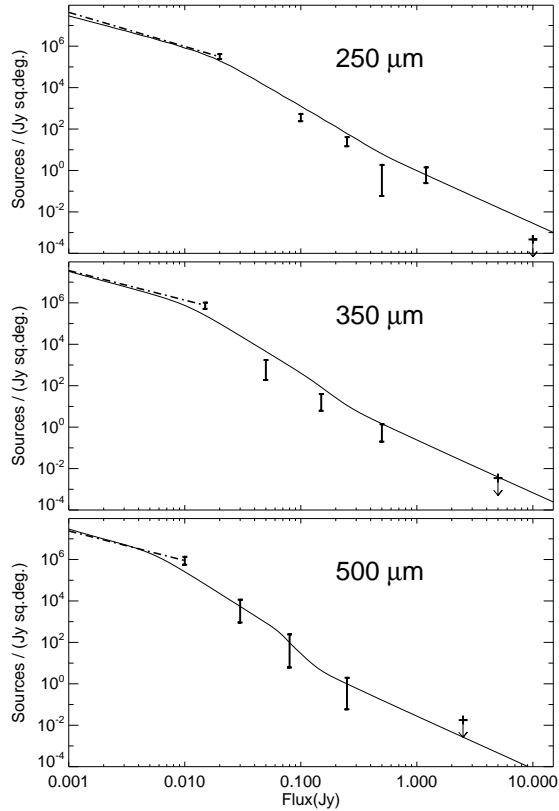}
  \caption{The BLAST differential source counts in Table
    \ref{tab:counts+FIRAS} at all three wavelengths are compared to
    the models of \citet{LDP04}.  The arrows at the right in each
    panel denote 99.9\% upper limits from the highest flux density nodes listed
    in Table~\ref{tab:counts+FIRAS}.  The dashed lines at the left
    connect the lowest node plotted to the 99.9\% confidence upper
    limit for an additional node at approximately $100\,\mu$Jy, also
    listed in Table~\ref{tab:counts+FIRAS}. The overall agreement is
    quite striking.  However, the model predicts more sources at 250
    and $350\,\mu$m than are observed and at all wavelengths the
    counts fall more steeply with flux than the model predicts.}
 \label{fig:comparewmodels}
\end{figure}

We have plotted BLAST differential source counts obtained with a
constraint from the total intensity of the CIB, which are listed in
Table \ref{tab:counts+FIRAS} in Figure \ref{fig:comparewmodels}.  The
curves in the figure are the models of \citet{LDP04}, obtained from
their web page\footnote{\tt
  http://www.ias.u-psud.fr/irgalaxies/Model/\#SourceCounts}, dated 5
Aug. 2008 ).  The models in this regime are largely the sum of two
components, starburst and quiescent galaxies, with starbursts
dominating in the middle of the figure and quiescent galaxies forming
the bulk of the flatter tail at high flux densities.  The overall
agreement is striking given the precision of the data, especially
noting that this is the first data-set available at these wavelengths
and the model has not been tuned to fit these new data.  However, the
model does overpredict (by a factor $\sim$10) the number of sources with flux
densities around 0.1--0.2$\,$Jy at 250 and $350\,\mu$m, and at all three
wavelengths the measured slope is steeper than predicted by about half
a unit.

\subsection{Comparison with other methods}

Standard techniques to estimate number counts are based on extracting
sources by finding peaks in a beam-correlated map. The measured flux
density of sources must be corrected for biases like flux-boosting
caused by applying a S/N threshold on noisy data with steeply falling
counts, as well as boosting due to confusion with fainter
sources. Derived number counts must also be corrected for
incompleteness and false identifications. In any highly confused maps
as for BLAST, these biases are large and strongly dependent on the
underlying counts. So, although it possible to account for some of the
effects \citep[e.g.][]{Coppin05,Austermann09b}, in the end one has to
carry out extensive simulations, which effectively reproduces the
forward-modelling of the pixel histogram which we describe in this
paper.  Moreover, counting objects above some rms level does not use
all of the information in the map.
\begin{figure*}
  \includegraphics[scale=0.3]{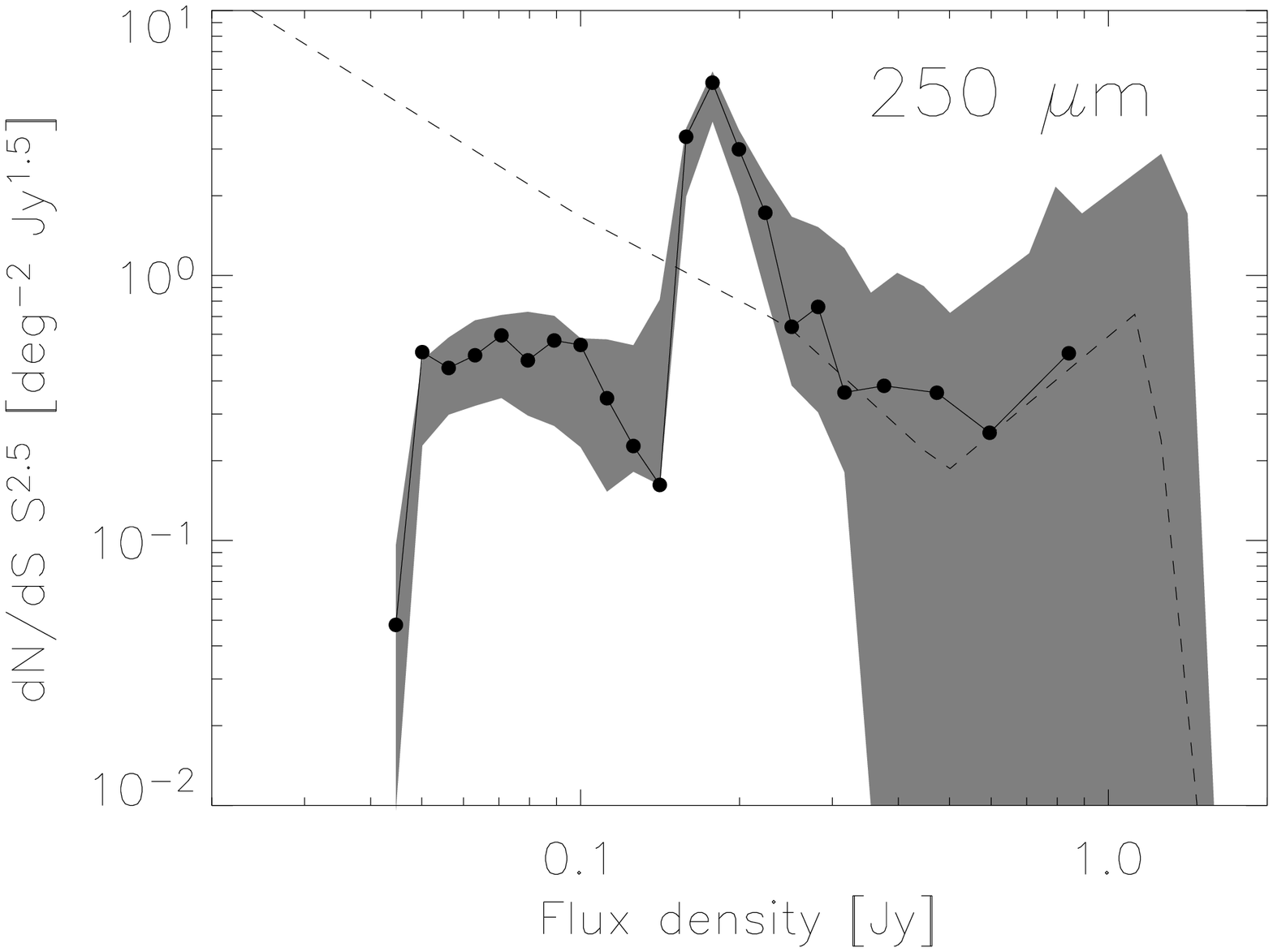}
  \includegraphics[scale=0.3]{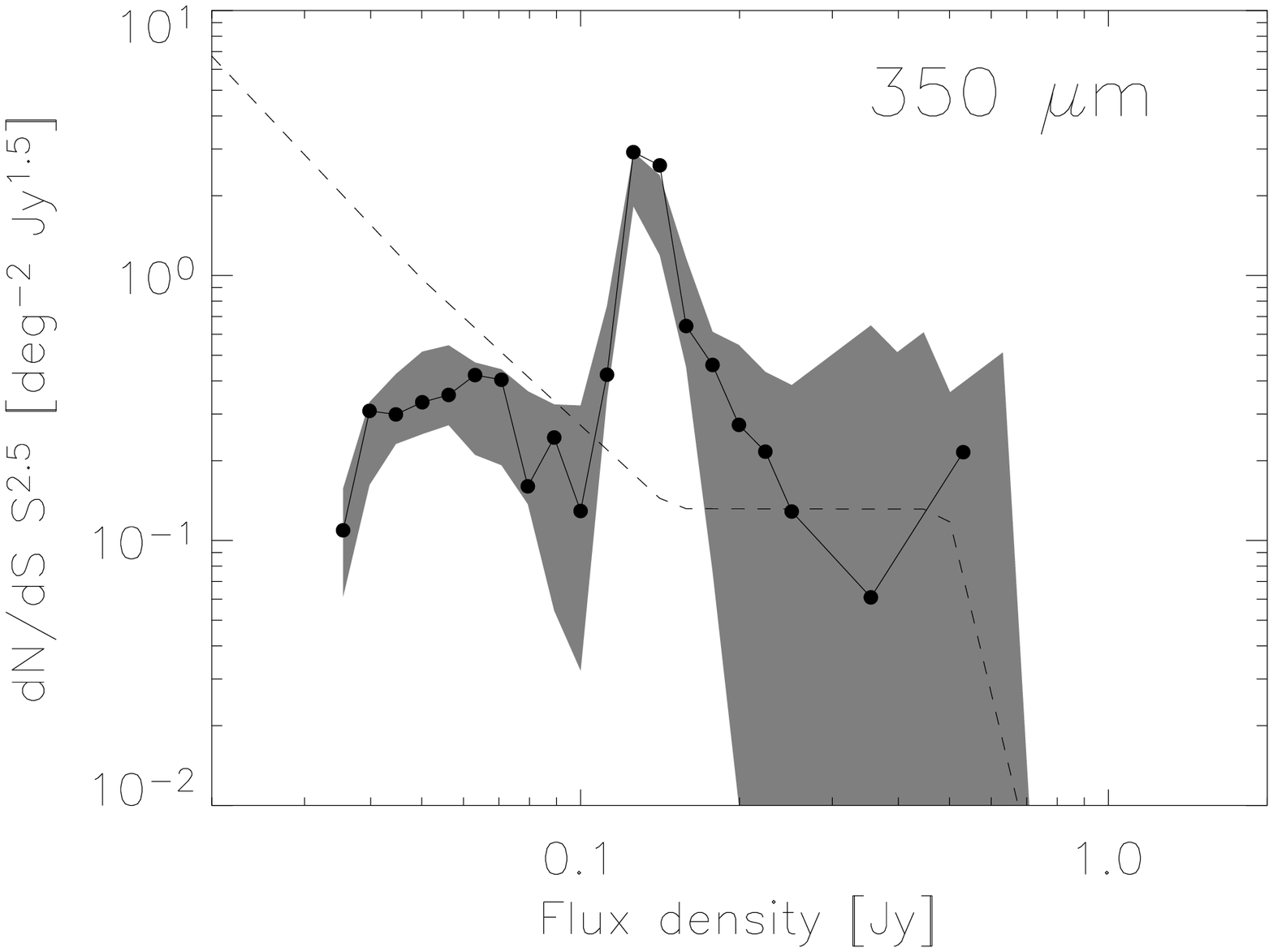}
  \includegraphics[scale=0.3]{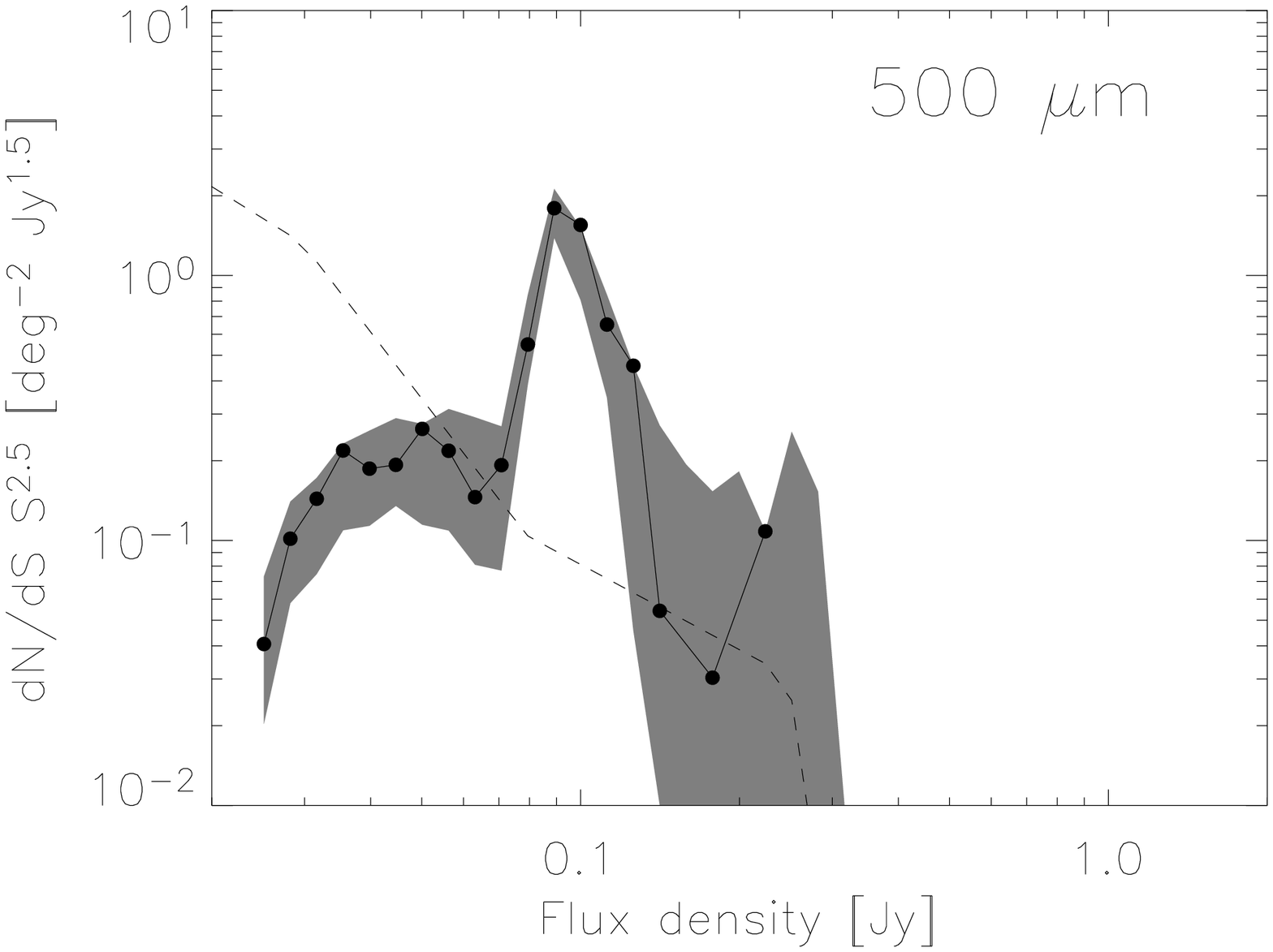}
  \caption{Comparison of Euclidean normalized differential counts with
    source catalogues at the three BLAST wavelengths. The dashed
    curves are best fit differential counts from the $P(D)$ analysis
    including FIRAS constraints. The solid curves are differential
    counts estimated by extracting 4-$\sigma$ sources from the BLAST
    maps and without applying any bias correction. The shaded region
    gives 95\% intervals for simulated 4-$\sigma$ catalogue counts
    distribution using multiple realizations of the BLAST maps. These
    shaded regions show what we {\it should\/} measure in the
    4-$\sigma$ catalogues if the dashed line is the correct underlying
    counts model. The negative bias at low flux densities is due to
    incompleteness and the `spike' at around $0.15\,$Jy at
    250\,$\micron$ is due to spurious detections and Eddington bias in
    the BGS-Wide region.  The excellent agreement between the
    catalogues estimated in BLAST maps (points and solid lines) and
    simulations (shaded regions) provides a satisfactory cross-check
    of our $P(D)$ analysis approach.  Using this method we recover
    counts which are consistent with the dashed lines.}
 \label{fig:scextractCompar}
\end{figure*}

Figure~\ref{fig:scextractCompar} shows the number counts estimated in
BLAST maps by counting objects but {\it without\/} applying any bias
correction, and compares it with counts estimated via the $P(D)$
analysis. One can see that biases are huge and even bigger than the
counts themselves at flux densities lower than about 0.1--$0.2\,$Jy at
all three wavelengths.

One could imagine an iterative approach for which the bias factor
correction would be derived from the debiased number counts, but one
would end up approaching the full $P(D)$ analysis. The method
presented in this paper efficiently provides unbiased estimates of the
counts over the full range in flux density (note in
Figure~\ref{fig:scextractCompar} the good match of the counts with
results from simulation using the best fit model).  It is therefore
highly recommended for confusion limited observations or for
multi-tier surveys where some of the layers suffer strong confusion.
Nevertheless, it may be that direct source estimation methods are a
little more efficient for the very brightest sources, for which the
probability of overlap is negligible.  That is because this makes
complete use of information from all pixels near the peaks.  In
practice this mild advantage for bright objects is probably of limited
use, since for model fitting one needs a single approach which spans
the full range of source brightnesses.  On the other hand source
extraction and recipes to correct for biases are still needed for
constructing catalogues and matching to other wavebands.  The point is
that it is the pixel information and not the source catalogue which
should be used to estimate the counts.

\subsection{Clustering of sources}
\label{sub:scClustering}

In the analysis presented in this paper, we have assumed that galaxies
are randomly and independently distributed over the sky. Nevertheless,
we know that all galaxies are clustered, and in fact significant
correlations have been found in the background of the BLAST maps
\citep{Viero09}, which are probably due to clustering on scales larger
than the BLAST beam.

The way that the probability distribution is modified depends strongly
on the angular scale being considered.  Clustering on scales much
larger than the beam will make the $P(D)$ distribution wider overall.
On the other hand, clustering around the beamsize (and smaller) will
distort the distribution in a way which depends much more on details
of the clustering model.

The effects can be computed for a given number counts model, provided
that all the $n$-point statistics of the source distribution are
known.  \citet{Barcons92} computed this for specific toy models of
clustering, while \citet{Toffolatti98} focussed on modelling of the
3-dimensional clustering of the sources.  \citet{Takeuchi04} describe
how to estimate the effects on the moments of the $P(D)$ distribution
by performing appropriate integrals over the angular 2-point
correlation function $w(\theta)$, as well as the higher $n$-point
correlations.  They show that for realistic source clustering the
effects of $w(\theta)$ dominate over those of higher-order
correlations, and lead to ${\sim}\,10$\% changes in the width of the
$P(D)$ histograms, together with a somewhat more extended tail.  The
reason that the impact of clustering is not larger is because most of
the effect comes from sources which are considerably fainter than the
confusion limit.  There are two consequences of this: these faint
sources have a relatively shallow number counts slope; and the
faintest sources are less strongly clustered than the brighter ones.
Simple scaling of the calculations of \cite{Takeuchi04} suggest that
the effects will be no larger for BLAST than for the other surveys
which they simulate.

We have carried out a determination of the clustering of BLAST sources
in Viero et al.~(2009), and fit this to models.  We show in Appendix~A
how to take a model for $P(k)$, derived from $w(\theta)$, and estimate
the effect of clustering on the $P(D)$ distribution.  In particular
the effect on the width of the distribution is relatively easy to
estimate, under a set of fairly reasonable assumptions.  We find that
the width of the distribution is increased by 13\%, 14\% and 20\% at 250,
350 and $500\,\mu$m, respectively. After re-convolving the map with
the beam kernel, the effect of clustering on the width of the
distribution becomes more important since the clusturing signal is
large scale, and hense the Poisson distribution get more reduced by
the convolution. In that case, we find that the width is increased by
28\%, 25\% and 30\%. These values are low enough compared with the
uncertainties that we are justified in neglecting the effects of
clustering on the counts; in particular, it seems that the effect is at most
of the order of one sigma for the fainter flux density
bins. However, for more precise estimates coming from future
data-sets, it may be that such clustering effects will need to be
fully considered.

\section{Conclusion}
\label{sec:conclusion}

We provide measurements of differential number counts at 250, 350, and
500\,$\micron$ from the statistical analysis of BLAST maps using a
maximum likelihood method based on 1-point statistics. We show that in
the SNR regime of BLAST and future surveys, this method is better
suited than counting individual sources, even when they are relatively
bright. This is because it naturally allows for the correction of
strong biases due to confusion and flux boosting.  This technique also
has the advantage of providing an unbiased estimate of the counts at
flux densities well below the limit at which sources can be detected
individually.  The method has been optimized to deal with
inhomogeneous noise across the map and filtering to suppress large
scale noise.

We measure the counts at a few flux nodes connected by power laws,
covering a wide range of fluxes, and perform careful analysis of the
resulting uncertainties using a Markov Chain Monte Carlo approach.  We
observe a very steep slope for the counts at intermediate flux densities
(approximately in the interval 0.02--0.5$\,$Jy) of about $-3.7$ at
250$\,\mu$m and $-4.5$ at 350 and 500$\,\mu$m, indicating strong
galaxy evolution.  We also detect a faint end break at all three
wavelengths at about $0.015\,$Jy.  Additionally we observe a change toward a
shallower slope at the bright end of the counts, particularly at
250$\,\micron$, consistent with an approach to the Euclidean regime.

The estimates and uncertainties we provide can be used for fitting to
specific models.  In addition the formalism presented in this paper
can also be applied directly to predict the histogram from
parametrized physical models like those derived in
\cite{Lagache03}. Comparison to the models of \cite{LDP04} show a
striking agreement given that the model has not been tuned to fit
these data, however we find fewer sources than expected by the model
at 250 and 350 $\mu$m and the slope of counts is steeper than expected
at all three wavelengths. Perhaps the contribution of quiescent
sources have been overestimated by the models.

We show that our method provides near optimal results when applied to
a map filtered with the beam kernel. Nevertheless, in the current
approach each map is treated independently, and consequently some of
the information contained in the combined data is not used. Since the
same sources are statistically detected at the three BLAST
wavelengths, but with amplitude ratios which depend on the intrinsic
spectra and redshifts, then one could use this additional information
to constrain more comprehensive models for the underlying sources.
The natural extension of our method would be the generalization to
three-dimensional histogram fitting of the form $P(D_1,D_2,D_3)$.  The
additional cross-band 1-point information brings additional
constraints on a combination of redshift evolution of the sources and
spectral energy distribution shapes.  Of course one could also imagine
an even more general fitting procedure which also uses the 2-point
statistics of the sort described in \citet{Viero09}, including the
cross-band clustering signals.

This paper provides the details required for developing the powerful
$P(D)$ technique to estimate counts from the much more extensive data
that will come from the SPIRE and PACS instruments on the {\em
  Herschel\/} satellite, as well as point sources in the {\em
  Planck\/} data.

\acknowledgments

We acknowledge the support of NASA through grant numbers NAG5-12785,
NAG5- 13301, and NNGO-6GI11G, the NSF Office of Polar Programs, the
Canadian Space Agency, the Natural Sciences and Engineering Research
Council (NSERC) of Canada, and the UK Science and Technology
Facilities Council (STFC). This research has been enabled by the use
of WestGrid computing resources.

\appendix
\section{Estimate of clustering effect}
\label{sec:appendix}

The expression for the variance of the pixel value distribution in the
presence of clustering is derived in Eq.~53 of \cite{Takeuchi04}, and
can be written as:
\begin{equation}
  \sigma^2 = \sigma_{\rm p}^2 + \sigma_{\rm c}^2.
\end{equation}
Here $\sigma_{\rm p}^2$ is the variance of pure Poisson fluctuations, and
$\sigma_{\rm c}^2$ the excess variance due to clustering:
\begin{eqnarray}
  \sigma_{\rm p}^2 & = & \int_{\Omega_b} \int_S S^2f(\bold{r})^2 n(S)
 dSd^2\bold{r}\label{eq:sigp}\\
  \sigma_{\rm c}^2 & = & \int_{\Omega_b} \int_{\Omega_b}
 \int_{S_1} \int_{S_2} S_1S_2 f(\bold{r_1}) f(\bold{r_2})
 n(S_1) n(S_2) w_2(\bold{r_1},\bold{r_2})
 dS_1dS_2d^2\bold{r_1}d^2\bold{r_2}\label{eq:sigc}.
\end{eqnarray}
$w_2(\bold{r_1},\bold{r_2})$ is the 2-dimensional 2-point
correlation function for positions $\bold{r_1}$ and
$\bold{r_2}$. Usually, the variance is evaluated up to a cutoff $x_{\rm c}$
in the observed signal given by $x=S/f(\bold{r})$. Thus the integrals
must be expressed in terms of $x_1$ and $x_2$, and as a consequence
the evaluation of the variance requires the computation of a 6-D
integral.

In our case, we can compute Eqs~\ref{eq:sigp} and \ref{eq:sigc}
up to a cutoff in source signal $S$ that we choose to be relatively high
(e.g.~5 times the noise rms). This is a very good approximation for
the computation of the map variance after extracting very bright
sources, since those are easily identified, and number counts are steep
enough that such bright sources do not contribute widely to the
variance.  Since the sky is isotropic and homogeneous, the 2-point
correlation function depends only on the angular separation $\theta =
|\bold{r_1} - \bold{r_2}|$, and then the variances can be expressed in
Fourier space:
\begin{equation}
  \sigma_{\rm c}^2 = \left(\int_S S~n(S) dS\right)^2 \int P(|\bold{k}|)
 {\bar{f}}(\bold{k})^2 d^2\bold{k}.
\end{equation}
Here $P(|\bold{k}|)$ is the 2-dimensional power spectrum of
clustering, and is equivalent to the Fourier transform of $w_2(\theta)$,
while $\bar{f}(\bold{k})$ is the Fourier transform of the beam function
$f(\bold{r})$. The excess rms of the $P(D)$ distribution due to
clustering (as given in \S~\ref{sub:scClustering}) can then be computed through
the following expression:
\begin{equation}
  r = {\sqrt{{\sigma_{\rm p}^2 + \sigma_{\rm c}^2} \over {\sigma_{\rm p}^2}}}.
\end{equation}

\bibliography{PofD_paper_X}

\end{document}